\documentclass[twocolumn,preprintnumbers,amsmath,amssymb,aps,prb,longbibliography]{revtex4-1}
\usepackage{graphicx}
\usepackage{color}
\usepackage{dcolumn}
\usepackage{bm}
\usepackage{hyperref}
\begin{document}

\title{Using Principal Component Analysis to Distinguish Different Dynamic Phases in Superconducting Vortex Matter}  
\author{
 C. J. O. Reichhardt$^{1}$,	
	D. McDermott$^{2}$, 
 and C. Reichhardt$^{1}$
} 
\affiliation{
$^1$Theoretical Division,  
Los Alamos National Laboratory, Los Alamos, New Mexico 87545 USA\\ 
$^2$X-Theoretical Design Division, Los Alamos National Laboratory, Los Alamos, New Mexico 87545 USA\\ 
}

\date{\today}
\begin{abstract}
Vortices in type-II superconductors driven over random disorder 
are known to exhibit a remarkable variety of
distinct nonequilibrium dynamical phases 
that arise due to the competition between
vortex-vortex interactions, the quenched disorder, and the drive. 
These include pinned states, elastic flows,
plastic or disordered flows, and dynamically reordered moving crystal
or moving smectic states.
The plastic flow phases can be particularly difficult to characterize
since the flows are strongly disordered.
Here we perform principal component analysis (PCA)
on the positions and velocities of
vortex matter moving over random disorder for different disorder strengths
and drives.
We find that PCA can distinguish the known
dynamic phases as well as or better than
previous measures based on transport signatures or topological defect
densities.
In addition, PCA recognizes distinct plastic flow regimes,
a slowly changing channel flow and a moving amorphous fluid flow,
that do not produce distinct
signatures in the standard measurements.
Our results suggest that this position and velocity based PCA
approach could
be used to characterize dynamic phases in
a broader class of systems that
exhibit depinning and nonequilibrium phase transitions.
\end{abstract}
\maketitle
    
\section{Introduction}

A wide variety of systems can be modeled effectively
as particles that, when driven over a random substrate,
can exhibit elastic or plastic depinning transitions as well as
distinct sliding states
\cite{Fisher98,Reichhardt17}.
Specific examples 
include vortices in type-II superconductors
\cite{Jensen88a,Bhattacharya93,Koshelev94,Higgins96,LeDoussal98,Balents98,Olson98a,Pardo98},
colloidal particles
\cite{Reichhardt02,Pertsinidis08,Tierno12a}, Wigner crystals
\cite{Cha94,Reichhardt01,Madathil23},
active matter \cite{Morin17,Sandor17a}, bubble and stripe
forming systems \cite{Reichhardt03a,Zhao13},
flowing emulsions \cite{LeBlay20},
granular matter \cite{Reichhardt12},
and magnetic skyrmions \cite{Iwasaki13,Reichhardt15}.
Related systems
include sliding charge density waves \cite{Bhattacharya87,Brazovskii04},
systems with interface depinning \cite{Leschhorn97},
dislocation depinning \cite{Moretti04,Zhou15},
sliding friction \cite{Vanossi12},
and fluid flow in disordered media \cite{Narayan94,Aussillous16}.
Depending on the strength of the disorder,
the particles may undergo elastic depinning and remain in a lattice
with the same neighbors \cite{Fisher98,DiScala12}.
Alternatively, they may exhibit
plastic depinning in which individual particles or small groups
of particles remain pinned while other particles flow past,
causing
the lattice to break apart into a
fluid-like state containing
a large number of topological defects
\cite{Jensen88a,Faleski96,Fisher98,Tonomura99b,Reichhardt02,Pertsinidis08,Fily10,Reichhardt17}.
For the plastically flowing system,
at higher drives the particles
can dynamically reorder into a lattice containing
a much smaller number of defects, such
as an
anisotropic moving crystal \cite{Koshelev94,LeDoussal98,Reichhardt15}
or moving smectic state
\cite{Moon96,Ryu96,Olson98a,Balents98,Pardo98,Kolton99,Fangohr01a,Reichhardt01}.
The pinning can continue to affect the fluctuations that occur in the
dynamically reordered state and can modify
the orientation of the topological defects that are still present.
The different dynamical phases are typically identified through features or
signatures in the transport or velocity-force
curves, such as the way in which the
velocity scales with the driving force just above depinning,
the specific shape of the full velocity-force curve, or the
presence of
peaks in the differential velocity-force curve
\cite{Jensen88,Bhattacharya93,Ryu96,Olson98a,Kolton99,Kokubo07,Fily10,Reichhardt17}.
The phases can also be characterized by
measuring the topological defect density.
Plastic flow states typically contain
numerous defects,
while the dynamically ordered phases contain a small number of defects that all
have their
Burgers vectors aligned in a particular direction.
The different dynamical phases can also be identified by analyzing
the noise characteristics of the velocity fluctuations,
which typically show a $1/f^\alpha$
signature in the plastic flow regime
\cite{Rabin98,Olson98,Olson98a,Okuma00,Reichhardt01}
and narrow band noise in the moving crystal or smectic flow
regimes
\cite{Tsuboi98,Kolton99,Togawa00,Olson98a,Bullard08,Madathil23}.

Particles moving over random disorder are
ideal systems in which to study universal features
of nonequilibrium phase transitions
\cite{Hinrichsen00,Sethna01,Corte08},
and there is now evidence that several of the dynamic phase transitions
in driven superconducting vortices fall
in the class of directed percolation 
\cite{Mangan08,Maegochi22a,Reichhardt23,Maegochi24}.
One of the biggest issues
in driven disordered systems and nonequilibrium phase transitions
is how to produce measures or
order parameters that can
distinguish between different phases,
particularly when the phases in question are disordered.
In systems that undergo
plastic depinning, there is evidence
suggesting that there could be multiple distinct plastic flow states.
Simulations show a transition between plastic flow states in which the
particles move along quasi-one-dimensional (q1D) channel structures and
states in which multiple moving channels begin to interact
\cite{GronbechJensen96,Fily10,Reichhardt22}.
Other work suggests that
the plastic flow can be characterized by how much of the
motion occurs in the direction transverse to the drive and that this
quantity changes when the nature of the plastic flow changes
\cite{Watson97,Reichhardt98,Mehta99,Bassler99,LeBlay20}.
Plastic depinning can also
be associated with large scale collective
motion or the translation of grain boundaries\cite{Cha98,Moretti09},
or can be connected to burst-like flow
\cite{Olson97,Bassler98,Reichhardt22}.
The flow states have also been characterized in terms of whether they
involve continuous or pulse-like motion
\cite{Faleski96}.
In general, there is no standardized
method for distinguishing different types of plastic flow.

The complexity of the driven dynamics in
systems with quenched disorder suggests that
machine learning (ML) methods could provide a fruitful characterization
approach.
In particular, principal component analysis (PCA) \cite{Abdi10,Shlens14}
can identify patterns in large data sets by reducing a high
dimensional feature vector into low dimensional order parameters.
PCA has been used to characterize a variety of complex systems
including biological systems \cite{McKinney06},
pattern forming states \cite{Bishop06},
and condensed matter systems \cite{Wang16,Carrasquilla17,Wetzel17,Hu17}.
PCA has successfully distinguished different phases and the
transitions between them
for equilibrium spin models \cite{Hu17} as well as
both equilibrium \cite{Jadrich18}
and nonequilibrium \cite{Jadrich18a}
off-lattice particle-based models.
It can also
capture
strongly heterogeneous nonequilibrium regimes for driven disks moving
over random disorder \cite{McDermott20} and
has been used to identify
distinct regimes of
motility-induced phase separation in active matter systems \cite{McDermott23}.

In this work, we consider superconducting vortices driven over
random quenched disorder of varied strength.
When the pinning strength is low, the vortices depin elastically and
form a moving crystal at higher drives,
while for higher pinning strengths, the depinning transition is
plastic, the vortices continuously exchange nearest neighbors,
and a dynamically reordered phase appears at high drives.
We construct ML feature vectors based on
information about both
the positions and the velocities of the particles in localized
neighborhoods,
and show that PCA can not only capture the known transitions
among the pinned, elastic, plastic, and dynamically ordered states,
but can also generate more pronounced signatures of these transitions
compared to standard measurement techniques such as the transport curves
or the topological defect density.
PCA additionally detects
distinct plastic flow regimes and provides order parameters for the
transitions between these regimes.
At lower drives, the plastic motion occurs via well-defined channels of
moving vortices flowing around two-dimensional islands of pinned
vortices, while at higher drives, the plastic flow behaves much more like
a moderately anisotropic liquid and both the channel structure and the
two-dimensional pinned islands are lost.
Our results suggest that PCA could be
a powerful tool for understanding the wider class of systems that
exhibit depinning, such as the onset of soliton motion
or different types of transitions from one- to two-dimensional flow
for particles moving over a periodic substrate,
depinning of domain walls, and avalanche systems \cite{Reichhardt17}.

\section{METHODS}
\subsection{Simulation}

We consider a two-dimensional system of size $L \times L$ containing
$N$ superconducting vortices
driven over a random substrate.
The system has periodic boundary conditions and we take $L=36\lambda$,
where $\lambda$ is the London penetration depth.
We measure all lengths in units of $\lambda$.
We shorten the system
size very slightly along the $x$ direction in order to accommodate a
triangular vortex lattice.
The vortex density is given by
$\rho = N/L^2$, and throughout this work we fix $N=1216$ and $\rho=1.0$.
The dynamics of vortex $i$ are obtained 
by numerically integrating the following overdamped equation of motion:
\begin{equation}
\eta \frac{{\bf R}_{i}}{dt}  =
{\bf F}^{vv}_{i} + {\bf F}^{vp}_i +  {\bf F}^{D} 
\end{equation}
where the damping term 
$\eta$ is set to unity.
The repulsive vortex-vortex interaction 
term is modeled as
${\bf F}_{i}^{vv} = \sum^{N}_{j\neq i}F_{0}K_{1}(R_{ij}/\lambda){\hat {\bf R}_{ij}}$,
where
$F_{0} = \phi^2_{0}/2\pi\mu_{0}\lambda^3$,
$\phi_0$ is the elementary flux quantum,
$\mu_{0}$ is the permittivity, $R_{ij} = |{\bf R}_{i} - {\bf R}_{j}|$
is the distance between particles $i$ and $j$,
$\hat {\bf R}_{ij}=({\bf R}_i-{\bf R}_j)/r_{ij}$, and $K_1$ is the Bessel
function of the first kind,
which decays exponentially at large $R$.
For computational efficiency,
we cut off the interaction for $R_{ij}/\lambda > 6.0$.

We model the random quenched disorder as
$N_{p}$ non-overlapping pinning sites, each of which consists of
an attractive parabolic potential well
with maximum range $r_{p}=0.3\lambda$ and a maximum strength of $F_{p}$.
The pinning force is
${\bf F}_i^{vp}=\sum_{k=1}^{N_p}(F_p/R_p)\Theta({\bf r}_{ik}^{(p)}-R_p)\hat {\bf r}_{ik}^{(p)}$, where
the distance between vortex $i$ and pin $k$ is
${\bf r}_{ik}^{(p)}={\bf r}_i-{\bf r}_k$,
$\hat {\bf r}_{ik}^{(p)}=({\bf r}_i-{\bf r}_k)/|{\bf r}_{ik}^{(p)}|$,
and $\Theta$ is the Heaviside step function.

We do not consider thermal effects in this work.
We initialize the vortices via a current annealing process, in which we
place the vortices into the sample in a triangular lattice, apply a current
${\bf F}^D=F_D {\bf {\hat x}}$
that is in the plastic flow regime, and gradually reduce the current
to zero, resulting in a distorted or heavily distorted vortex lattice that has
accommodated itself to the quenched disorder.
This procedure gives us an initial zero drive state that is disordered, so
that there are no ordered state transients below depinning
of the type
that appear in our earliest study of this system \cite{Olson98a}.
If we instead use a thermal annealing procedure to produce the initial
pinning-accommodated state, the results are unchanged.
Beginning from the initialized configuration,
we collect data while sweeping the driving
current from $F_D=0$
to a maximum value of $F_D=2F_p$ in increments of
$\Delta F_D=0.005F_p$,
spending $1 \times 10^5$ simulation time steps at each current increment.
This sweep rate is slow enough to avoid
transient effects at individual currents
and provides us with high resolution velocity-force data in the
different plastic and elastic flow regimes.
We write out the data every 1000 simulation time steps,
so that we obtain $N_f=100$ frames of data for every value of the
current.
Unless otherwise indicated, all results shown in this work have been
averaged over five realizations of disorder.

\subsection{Principal Component Analysis}

In previous studies of dynamical phase transitions
in vortex systems, the phases were identified based on
features in the velocity-force curves, topological defect
density, and structure factor, or through a
time series analysis of the
velocity fluctuations.
Here, we consider some of these same measures, including
the average velocity in the driving direction,
$V=\langle \sum_i^N v_i \cdot {\bf {\hat x}}\rangle$
and the percentage of sixfold coordinated vortices,
$p_6=\langle \sum_i^N \delta(z_i-6)\rangle$,
where $z_i$ is the coordination number of vortex $i$ obtained from
a Voronoi tesselation
and the averages are taken over individual current increments.
For a perfect triangular lattice, $p_6=1.0$.
In addition, we perform
a principal component analysis (PCA)~\cite{Abdi10} on feature
vectors extracted from the vortex configurations and velocity components.
PCA identifies the directions of maximum variance in the space defined
by the feature vectors.
The PCA approach we employ is based on that used
for the analysis of off-lattice nonequilibrium systems
\cite{Jadrich18,McDermott20,McDermott23}.
We tested the
same type of position-based (PB) feature vectors used in previous
work, and also developed position-and-velocity-based (PVB) feature
vectors specifically tailored for exploring the plastic flow states.

To construct a PB feature vector, we inspect individual frames of
the simulation data and select $N_p$ probe vortices at random.
For each probe vortex $i$,
we compute the distance
$r_{ij} = |{\bf r}_{i}-{\bf r}_j|$
to all of the other vortices, normalize this by the average
lattice constant $a_0=\sqrt{N/L^2}$,
and sort these distances in ascending
order. To speed the analysis, we drop vortices that
are at a distance greater than
$r_{ij}=8\lambda$ before performing the sort.
We then take the $n$ smallest distances and place them into an array
in increasing order of distance,
\begin{equation}
  {\bf f}_i^{\rm raw} = [r_{i0},r_{i1},r_{i2}, ..., r_{ij}, ..., r_{in}] 
 \label{eq:feature}
\end{equation}
where $r_{i0} < r_{i1} <  ... < r_{in}$.
We repeat this procedure for all $N_p$ probe vortices, and obtain
the average distance to neighbor $n$ in frame $m$ of the movie as
$\bar{r}_{n}=N_p^{-1}\sum_i^{N_p} r_{in}$.
This gives us our final feature vector for frame $m$,
\begin{equation}
  {\bf f}_m = [\bar{r}_{m0},\bar{r}_{m1},\bar{r}_{m2}, ..., \bar{r}_{mj}, ..., \bar{r}_{mn}] .
\end{equation}
By processing each frame in the entire simulation data set,
we accumulate a matrix ${\bf M}$ in which each row is the vector
${\bf f}_m$ from frame $m$. There are a total of $2\times 10^4$ rows
in matrix ${\bf M}$, since there are $N_f=100$ frames per current and there are
200 current values in the velocity sweep.
In previous work on particles with short range interactions, it was
necessary to prewhiten the feature vector against a PCA analysis of
an ideal gas in order to remove spurious geometric information from
the feature vector \cite{Jadrich18, McDermott20, McDermott23}.
We find that prewhitening is not necessary for the vortex system
since the relatively long range interactions between vortices produces
a hyperuniform or nearly hyperuniform distorted triangular lattice.
Here, the hyperuniformity simply indicates that the average vortex
density is nearly constant across the entire sample. As a result,
spurious geometric signals caused by clumping of the particles do not
occur for the vortices.
We tested a variety of methods for selecting the value of $n$, and
obtained the cleanest results when we worked with
the full set of
consecutive neighbor distances without skipping any
consecutive distances.
For the results in this paper,
we use
$n=144$
and $N_p=50$.

To construct a PVB feature vector, we first perform the steps described
above to identify the $n$ closest neighbors of a given probe vortex,
which we will refer to as subset $\mathcal{N}$.
We sort the distances from the probe vortex to the vortices in
$\mathcal{N}$ into ascending order
as above and use these distances as the first $n$
entries of the raw feature vector ${\bf f}_i^{\rm raw}$.
Working {\it only} with the vortices in $\mathcal{N}$, we
separately sort the absolute $x$ direction velocities
$v_{x,i}=|v_i \cdot{\bf {\hat x}}|$ normalized by $F_D$
in ascending order and use them as the second $n$ entries in
${\bf f}_i^{\rm raw}$.
We then 
separately sort the absolute $y$ direction velocities
$v_{y,i}=|v_i \cdot{\bf {\hat y}}|$ 
in ascending order and use them as the third $n$ entries in
${\bf f}_i^{\rm raw}$ to obtain
\begin{equation}
  \begin{split}
  {\bf f}_i^{\rm raw} = & [r_{i0},r_{i1},r_{i2}, ..., r_{ij}, ..., r_{in},\\
  & v_{x,i0},v_{x,i1},v_{x,i2}, ..., v_{x,ij}, ..., v_{x,in}, \\
  & v_{y,i0},v_{y,i1},v_{y,i2}, ..., v_{y,ij}, ..., v_{y,in}]
  \end{split}
 \label{eq:feature}
\end{equation}
where $r_{i0} < r_{i1} <  ... < r_{in}$,
$v_{x,i0} < v_{x,i1} <  ... < v_{x,in}$,
and
$v_{y,i0} < v_{y,i1} <  ... < v_{y,in}$.
It is important to note that the vortices in $\mathcal{N}$ are
sorted three separate times while being used to construct
the feature vector, so that the neighbor with the smallest
distance $r_{i0}$ can be {\it different} from the neighbor with the smallest
$x$ velocity $v_{x,i0}$, and both of these can be different from
the neighbor with the smallest $y$ velocity $v_{y,i0}$. The neighborhood
$\mathcal{N}$ is defined based on the position vectors alone and is
not redefined when working with the velocity vectors.
To find the final feature vector, we average each entry of the feature
vector over all $N_p$ probe particles, giving
\begin{equation}
  \begin{split}
  {\bf f}_m = & [\bar{r}_{m0},\bar{r}_{m1},\bar{r}_{m2}, ..., \bar{r}_{mj}, ..., \bar{r}_{mn},\\
    & \bar{v}_{x,m0},\bar{v}_{x,m1},\bar{v}_{x,m2}, ..., \bar{v}_{x,mj}, ..., \bar{v}_{x,mn},\\
    & \bar{v}_{y,m0},\bar{v}_{y,m1},\bar{v}_{y,m2}, ..., \bar{v}_{y,mj}, ..., \bar{v}_{y,mn}]
  \end{split}
\end{equation}
where
$\bar{v}_{x,n}=N_p^{-1}\sum_i^{N_p} v_{x,in}$ and
$\bar{v}_{y,n}=N_p^{-1}\sum_i^{N_p} v_{y,in}$.
As before, we create a matrix ${\bf M}$ in which each row is the
vector ${\bf f}_m$ from frame $m$.
By constructing the PVB feature vector in this way, we capture information
about the average deformations in a localized area
of size $\mathcal{N}$ of the vortex lattice that are produced by
velocity differences among the vortices. As we show below, this
enables us to probe the detailed structure of the plastic vortex flow
based on individual frames of the simulation data without requiring
extremely high resolution imaging of the vortex trajectories.

Regardless of whether we are working with a PB or PVB feature vector,
we use standard PCA techniques to compute the
orthogonal transformation matrix
${\bf W}$ mapping our feature vectors of length $m$ to the principal components
based on data from the entire simulated velocity-force sweep for
a particular value of $F_p$,
 \begin{equation}
   {\bf p} = {\bf W}{\bf M}.
 \end{equation}
Here ${\bf W} \equiv [{\bf w}_i, ..., {\bf w}_m]^T$, where the unit
vectors ${\bf w}_i$ define the directions of the principal components.
To construct order parameters for
a current $F_D^i$, we project the individual feature vectors obtained
for that current along the principal component directions and average the
resulting values:
\begin{equation}
  P_\alpha^i=N_f^{-1}\sum_{k \in F_D^i}{\bf w}^T_\alpha{\bf f}_k/\sqrt{\lambda_\alpha} .
\end{equation}
Here the normalization factor is the square root of $\lambda_\alpha$,
the eigenvalue associated with principal component $\alpha$.
For PB feature vectors we use $\alpha=\{1, 2\}$, and
for PVB feature vectors we use $\alpha=\{1, 2, 3\}$.
Note that the eigenvectors ${\bf w}_i$ are only defined to within a sign
and remain eigenvectors if they are reflected across the origin.
As a result, the order parameters are also only defined to within a sign,
and although there is significance to the order parameter passing through
zero, since it means that the
effective dimensionality of the data has been
reduced by one,
there is no special significance to the fact that the order parameter
is positive or negative overall.
We note that the results of the PCA analysis are nearly unchanged
if we replace the absolute $x$ direction velocities
$|v_i \cdot {\bf {\hat x}}|$
in the PVB feature vector with the velocity magnitude $|v_i|$, since
for the system considered here, nearly the entire vortex velocity component
is oriented along the $x$ direction.

\section{Results}

\begin{figure}
\includegraphics[width=\columnwidth]{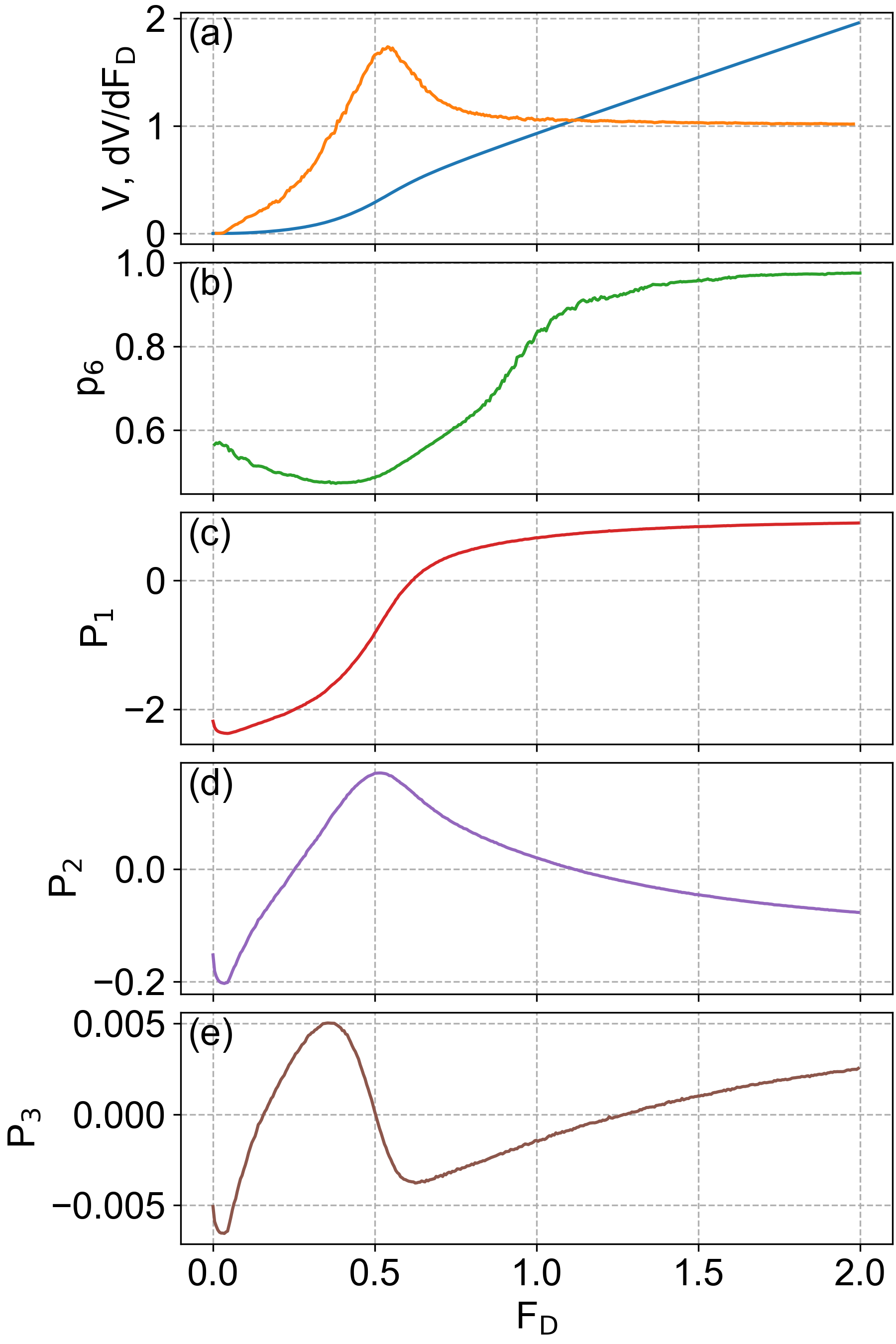}
\caption{(a) Velocity $V$ (blue) and $dV/dF_D$ (orange) vs driving force $F_D$ 
for vortices driven over random disorder in a sample
with pinning force $F_{p} = 1.0$.
There is a pinned phase, a plastic flow phase,
and a dynamically reordered phase.
(b) Corresponding fraction of sixfold coordinated vortices $p_6$ vs $F_D$.
(c-e) Corresponding PVB principal components $P_n$ vs $F_D$.
(c) $P_1$. (d) $P_2$. (e) $P_3$.}
\label{fig:1}
\end{figure}

In Fig.~\ref{fig:1}(a), we plot the $V$ and
$dV/dF_{D}$ versus $F_{D}$ curves for
a driven vortex system with $F_{p} = 1.0$,
where pinned, plastic, and dynamically reordered states occur
as a function of increasing drive.
The derivative
$dV/dF_{D}$ increases from zero at depinning and reaches a
peak near $F_{D} = 0.54$, then decreases more slowly
and approaches $dV/dF_D=1.0$ for $F_{D} > 1.25$.
In previous work, the increase and peak in $dV/dF_{D}$
have been associated with plastic depinning and flow, while
the saturation to
$dV/dF_D=1.0$ at higher drives
has been connected to a dynamical reordering transition
\cite{Reichhardt17}.
Figure~\ref{fig:1}(b) shows the corresponding
fraction of sixfold coordinated vortices $p_6$
versus $F_{D}$.
The current annealing procedure results in an initial pinned disordered
glass state with
$p_{6} = 0.59$ at low drives,
so that a fraction $1-p_6=0.41$ of the vortices are
associated with topological defects in the lattice. As $F_D$ increases
$p_{6}$ initially decreases and reaches a minimum of $p_6 \approx 0.5$
near $F_{D} = 0.45$.
At the minimum in $p_6$, $V$ falls well below the expected pin-free value
of $V=F_D$, and the system contains
a combination of pinned and slowly moving vortices.
For $F_{D} > 0.45$, $p_6$ increases as the density of topological defects
decreases before reaching a plateau
near $F_{D} = 1.25$. At the highest drives we find
$p_{6} = 0.96$, indicative of the appearance of a dynamically ordered
moving smectic state.

\begin{figure}
\includegraphics[width=\columnwidth]{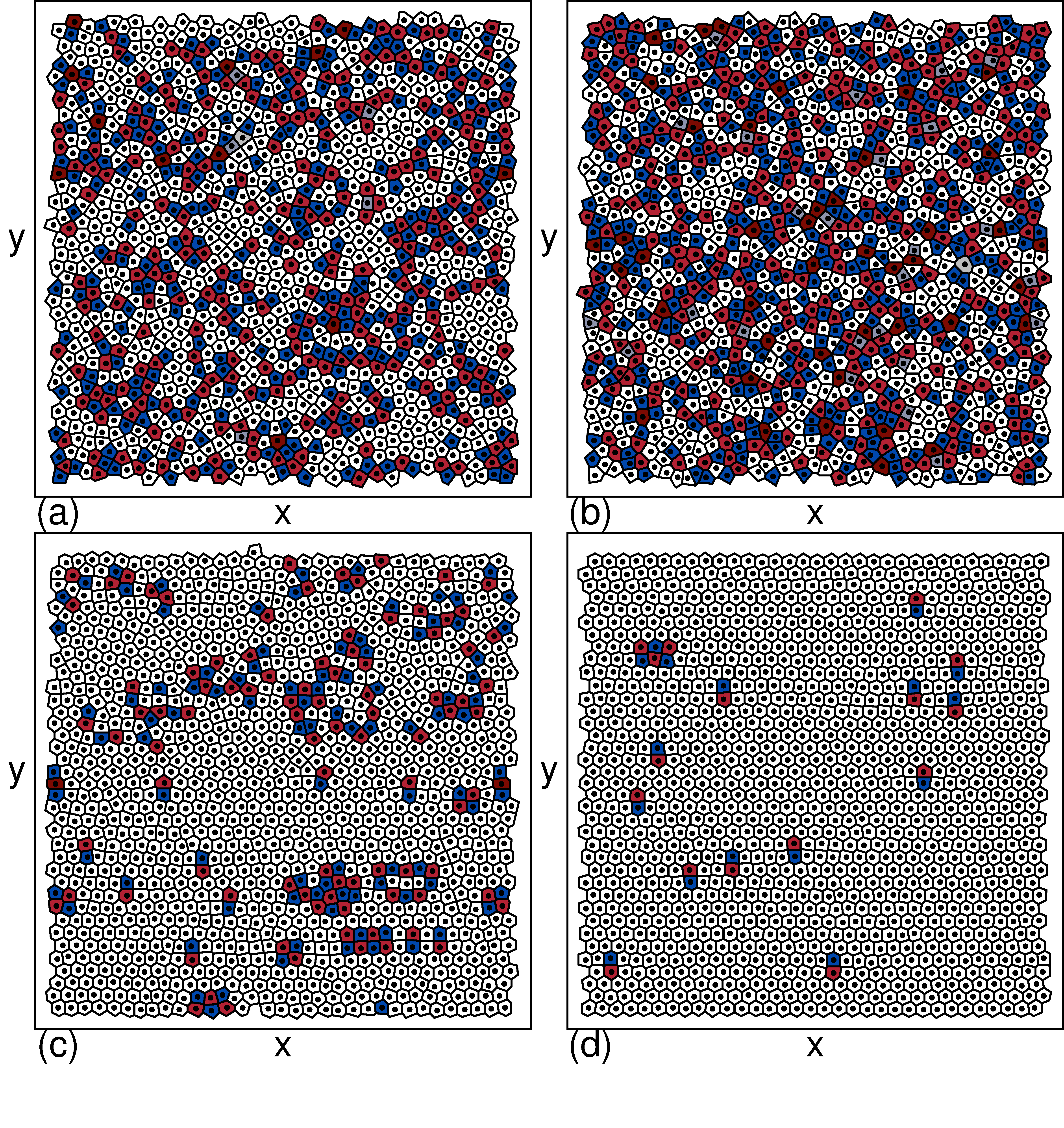}
\caption{Voronoi constructions of snapshots of vortex positions for
  one quenched disorder realization of the
  system in Fig.~\ref{fig:1} with $F_p=1.0$ at
  (a) $F_D=0.0$ in a pinned amorphous state,
  (b) $F_D=0.5$ in the plastic flow state,
  (c) $F_D=1.0$ in a partially disordered state,
  and $F_D=2.0$ in the moving smectic state.
  Dots indicate the vortex locations and the polygons are colored according
  to coordination number: $z_i=5$ (blue), $z_i=6$ (white), 
  $z_i=7$ (light red), and all other non-sixfold coordination numbers (all
  remaining colors).
	}
\label{fig:2}
\end{figure}

In Fig.~\ref{fig:2}, we show Voronoi construction images at selected
drives for the $F_p=1.0$ system from Fig.~\ref{fig:1}.
In the pinned phase at $F_D=0.0$, 
Fig.~\ref{fig:2}(a) indicates that many topological defects
are present and the system forms a pinned glass.
At $F_D=0.5$ in Fig.~\ref{fig:2}(b), the vortices are flowing plastically
and are close to their maximally disordered state.
The lattice partially reorders for $F_D=1.0$ in Fig.~\ref{fig:2}(c), where
there are large regions of sixfold-coordinated vortices
as well as smaller regions containing numerous topological defects.
In Fig.~\ref{fig:2}(d) at $F_{D} = 2.0$,
the system has formed a moving smectic in which the vortices
travel in q1D channels that can
slip past one another.
A small number of paired fivefold and sevenfold coordinated defects are
present, each with its Burgers vector aligned in the driving direction.

The PVB principal component derived order parameters
$P_1$, $P_2$, and $P_3$ appear in
Fig.~\ref{fig:1}(c,d,e) as a function of $F_D$.
Here the feature vectors incorporate information about
the vortex positions, velocity parallel to the drive, and velocity
perpendicular to the drive.
Figure~\ref{fig:1}(c)
shows that although $P_{1}$ is broadly similar to $p_{6}$ in shape, it has
distinctive features.
A minimum in $P_1$ appears at the depinning transition when the vortices
begin to move, but there is
no similar feature in $p_6$.
There is a continuous increase in $P_1$ for drives above depinning and
a zero crossing at $F_D=0.62$.
The curve begins to saturate
near $F_{D} = 0.7$, which is a lower
$F_{D}$ value than the point at which $p_6$ begins to saturate.
The plot of $P_2$ versus $F_D$ in
Fig.~\ref{fig:1}(d) also shows a dip near the depinning transition and
a zero crossing at $F_D=0.23$ in the plastic flow state,
as well as a peak
at $F_D=0.52$, which is slightly smaller than the $F_D$ value at which
$dV/dF_{D}$ reaches its maximum value.
Although $P_2$ has no sharp feature at the dynamic reordering
transition that occurs at higher drives, it does have a second
zero crossing from positive to negative near $F_{D}  = 1.25$.
In Fig.~\ref{fig:1}(e), the $P_{3}$ versus $F_D$ curve
also shows a dip feature near depinning and has multiple zero crossings.
There is a local maximum near $F_{D} = 0.36$, a drive that
falls well below the drive at which
$dV/dF_{D}$ peaks.
There is also a local minimum near $F_D=0.62$,
which falls above the peak in $dV/dF_{p}$ but matches the
zero crossing of $P_1$.
We will show below that the zero crossing feature in $P_1$ and
the local minimum in $P_3$
at $F_{D} = 0.62$ mark the separation between
two different types of plastic flow phases.
At higher drives, $P_{3}$ crosses zero
again near $F_{D} = 1.25$,
and continues to rise inside the dynamically reordered smectic state.

Based on our comparison of the order parameters
$P_{n}$ and the features in $dV/F_{D}$ and $P_{6}$
as a function of drive, we find that the PVB PCA analysis
can successfully
capture changes that occur in the dynamics and structure
of the driven vortices.
We next consider whether additional information can be extracted
from the features in $P_n$,
such as whether the zero crossings, peaks, and dips in $P_2$ and $P_3$ are
detecting additional structured phases
within the plastic flow regime.
Below we demonstrate that the plastic flow can be broken into what
we call non-ergodic plastic flow, where some vortices are pinned permanently
and some regions of the sample are inaccessible to the motion,
and ergodic plastic flow,
where some vortices can be temporarily pinned but all vortices take part
in the motion and all regions of the sample can be accessed by the flow.
The PVB PCA analysis
also provides
evidence that there are
additional subphases within the non-ergodic and
ergodic plastic flow regimes.

\begin{figure}
\includegraphics[width=\columnwidth]{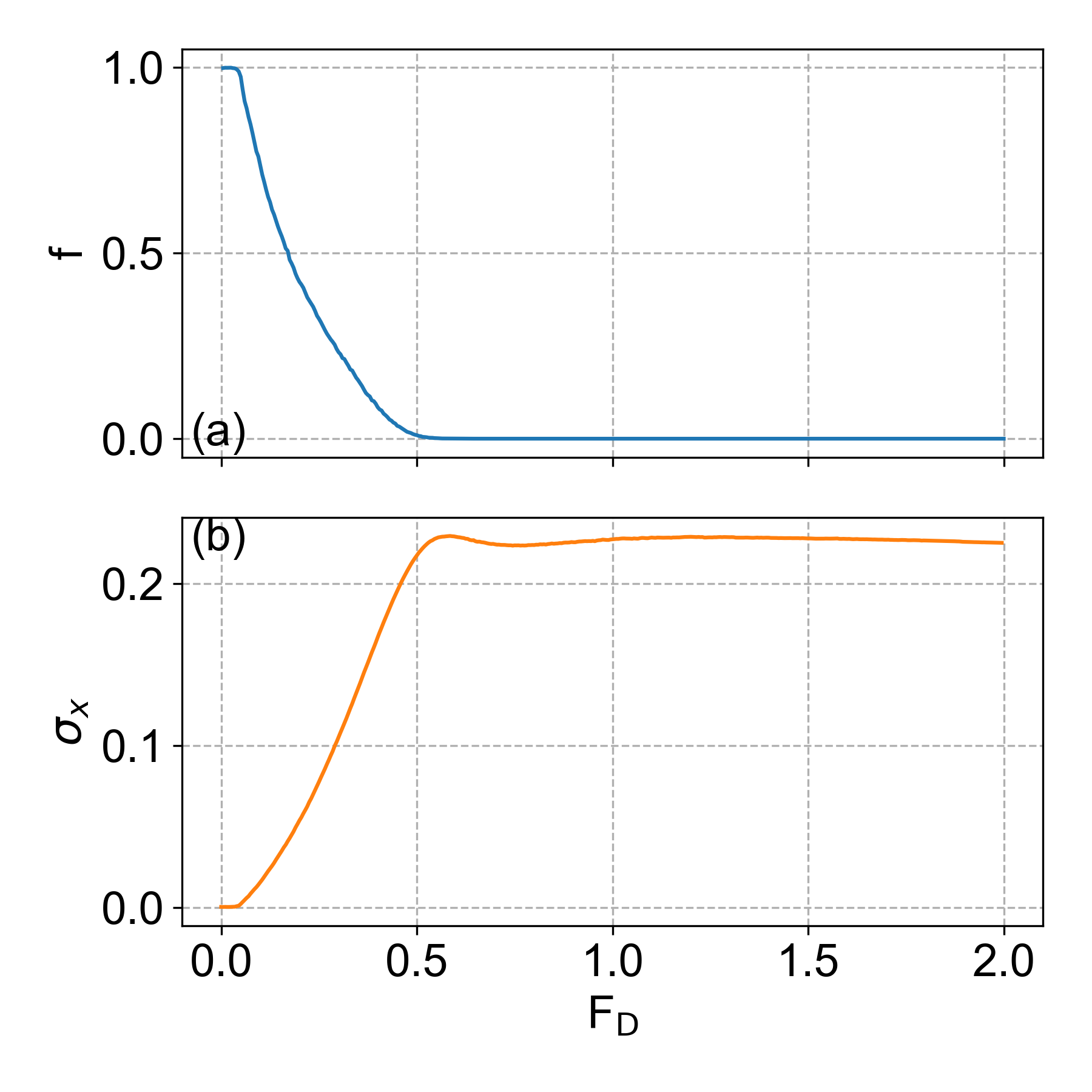}
\caption{(a) The fraction $f$ of permanently pinned vortices
  vs $F_{D}$ for the sample from Fig.~\ref{fig:1} with $F_p=1.0$.
  (b) $\sigma_x$, the standard deviation of
  the $x$ velocity $V$, vs $F_{D}$ for the same system.
}
\label{fig:3}
\end{figure}

In Fig.~\ref{fig:3}(a,b), we plot the fraction $f$ of permanently
pinned vortices versus and the standard deviation $\sigma_x$ of the
instantaneous $x$ direction velocity $V$ versus $F_D$ for the $F_p=1.0$ sample
from Fig.~\ref{fig:1}.
We post-process the value of $f$ using the vortex position information that
was written out every 1000 simulation time steps, and define a vortex to
be permanently pinned if it moved a distance equal to
less than the pinning radius between frames and also has an instantaneous
velocity that is smaller than $\Delta F_D$ (to avoid stroboscopic effects).
This gives
$f=N^{-1}\sum_i^N \Theta(r_p-|{\bf R}_i(t_n)-{\bf R}_i(t_{n-1})|)$,
where $t_n$ is the time at which frame $n$ was obtained and
we average $f$ over the second half of the frames belonging to each current.
The value of $\sigma_x$ is calculated during the simulation and is
given by
$\sigma_x=\sqrt{\sum_i^N(v_x^i-\mu_x)^2/(N-1)}$
where the mean velocity $\mu_x=N^{-1}\sum_i^N v_x^i$
and the
instantaneous
$x$ velocity of vortex $i$ is $v_x^i={\bf v}_i \cdot {\bf{\hat x}}$.
At $F_D=0$, $f=1$ in Fig.~\ref{fig:3}(a). The fraction of permanently pinned
vortices reaches zero at $F_D\approx 0.6$,
and in Fig.~\ref{fig:3}(b), $\sigma_x$
reaches its maximum value slightly earlier at $F_D=0.58$,
passes through a shallow minimum centered near $F_D=0.74$,
and then saturates above $F_D=F_p$. Due to the small density of
dislocations that remain inside the dynamically reordered smectic
state, the velocity distribution function remains slightly bimodal.
The point where $f$ reaches zero matches the zero crossing
of $P_1$ in Fig.~\ref{fig:1}(c)
and the local minimum of $P_3$ in Fig.~\ref{fig:1}(e),
but falls above the peak in $dV/dF_D$
in Fig.~\ref{fig:1}(a).
We note that even though all of the vortices participate in the motion
above $F_D\approx 0.6$,
the system remains topologically disordered
for $0.6 < F_{D} < 1.0$, and in this regime some of the vortices can
still be pinned temporarily.
We argue that the zero crossing point of $P_1$ is associated
with a transition or crossover
from what we call non-ergodic plastic flow, where some vortices never depin,
to ergodic plastic flow,
where all of the vortices participate in the motion over time.
There are no sharp features in $f$ or $\sigma_x$ at higher drives.

\begin{figure}
\includegraphics[width=\columnwidth]{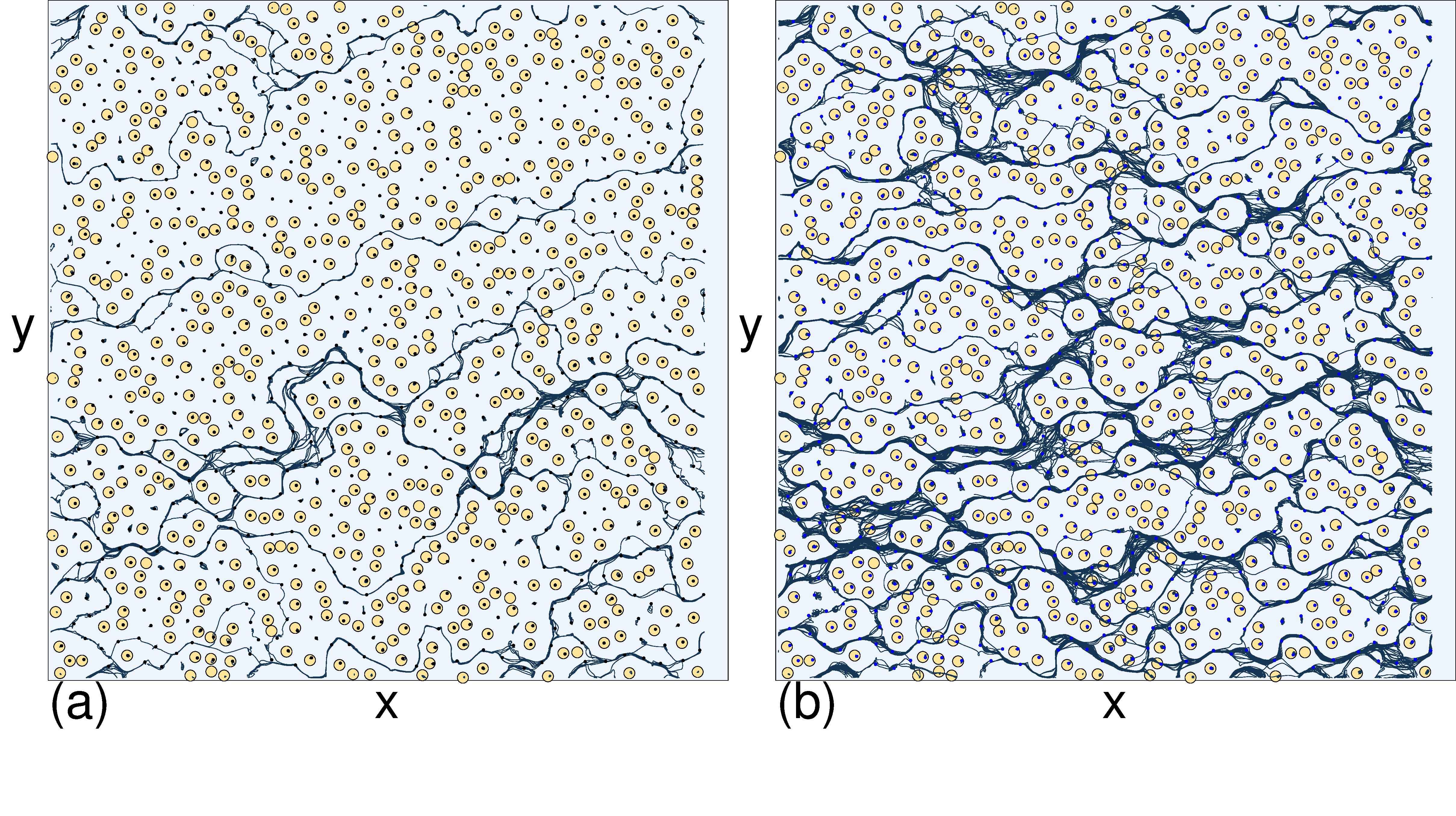}
\caption{Images of pinning sites (open circles),
vortex positions (dots), and vortex trajectories (lines) over a fixed
period of time
for the $F_p=1.0$ system from Fig.~\ref{fig:1}.
(a) At $F_{D} = 0.125$, a small number of very well defined flow channels
are present.
(b) At $F_{D} = 0.23$, the channel density has increased and adjacent
channels can interact, but there remain regions in the sample where
flow never occurs.
}
\label{fig:4}
\end{figure}

We image the vortex trajectories and the differences in the trajectories
over time in order to 
get an idea of how the motion changes
with increasing drive within the plastic flow regime.
The $F_p=1.0$ system from Fig.~\ref{fig:1} has sufficiently strong pinning
that the flow occurs in channels.
For $F_{D} < 0.25$, the channels are very well defined and their structure
is determined by the details of the quenched disorder.
In Fig.~\ref{fig:4}(a), we show the pinning sites, vortex locations,
and vortex trajectories for
$F_{D} = 0.125$ where the flow occurs in a small number of
well-defined channels.
At $F_D=0.23$ in Fig.~\ref{fig:4}(b),
the number of flowing channels has increased and there can be exchange
among vortices in adjacent channels, but there are still permanently pinned
vortices present inside regions of the sample where flow never
occurs.
If we image the trajectories over a longer period of time,
the vortices continue to flow in the same channels shown in 
the figure, indicating that the flow is
non-ergodic and does not visit the entire sample.
The drives illustrated in Fig.~\ref{fig:4} fall
below the zero crossing of $P_1$ in Fig.~\ref{fig:1}(c), below the
first zero crossing of $P_2$ in Fig.~\ref{fig:1}(d), and below
the local peak in $P_3$ in Fig.~\ref{fig:1}(e).

For higher drives,
the trajectories become increasingly smeared as the channels lose their
distinct nature, and plots of the type presented in Fig.~\ref{fig:4}
are not useful as they show only an indistinct tangle of trajectories.
Thus, to visualize the flow we first project the trajectories
from a fixed period of time onto a fine
discrete grid in order to
obtain a two-dimensional height field ${\bf H}$
showing the number of times a location in space was
visited by a moving vortex.
We perform the trajectory mapping using Bresenham's line algorithm.
The vortices must be moving in order to be
recorded on the height field; static vortices are not counted.
We repeat this for two well-separated
time intervals $T_1$ and $T_2$, and then take the difference
${\bf D}_h={\bf H}(T_2)-{\bf H}(T_1)$
of the resulting trajectory height
fields in order to determine whether the
channels of flow are fixed and thus non-ergodic, or are changing and thus
ergodic.
In regions of space
where no flow occurs, there are no trajectories and the difference
of the trajectory height fields is zero.
In regions of space
where a fixed flow channel is present, the same flow occurs
during both sampling times and the difference
of the trajectory height
fields is again zero. Only regions of space where the flow is not the same
during the two sampling time periods will have finite values in the
trajectory height field difference image.
We then color the height field
difference image with a diverging color map so that regions of
zero value are white.
Under this measure, a completely pinned system
is white, and a uniformly moving crystal is also white.
Non-ergodic flow regimes will have some white spatial regions,
and ergodic regimes will have no white regions.

\begin{figure}
\includegraphics[width=\columnwidth]{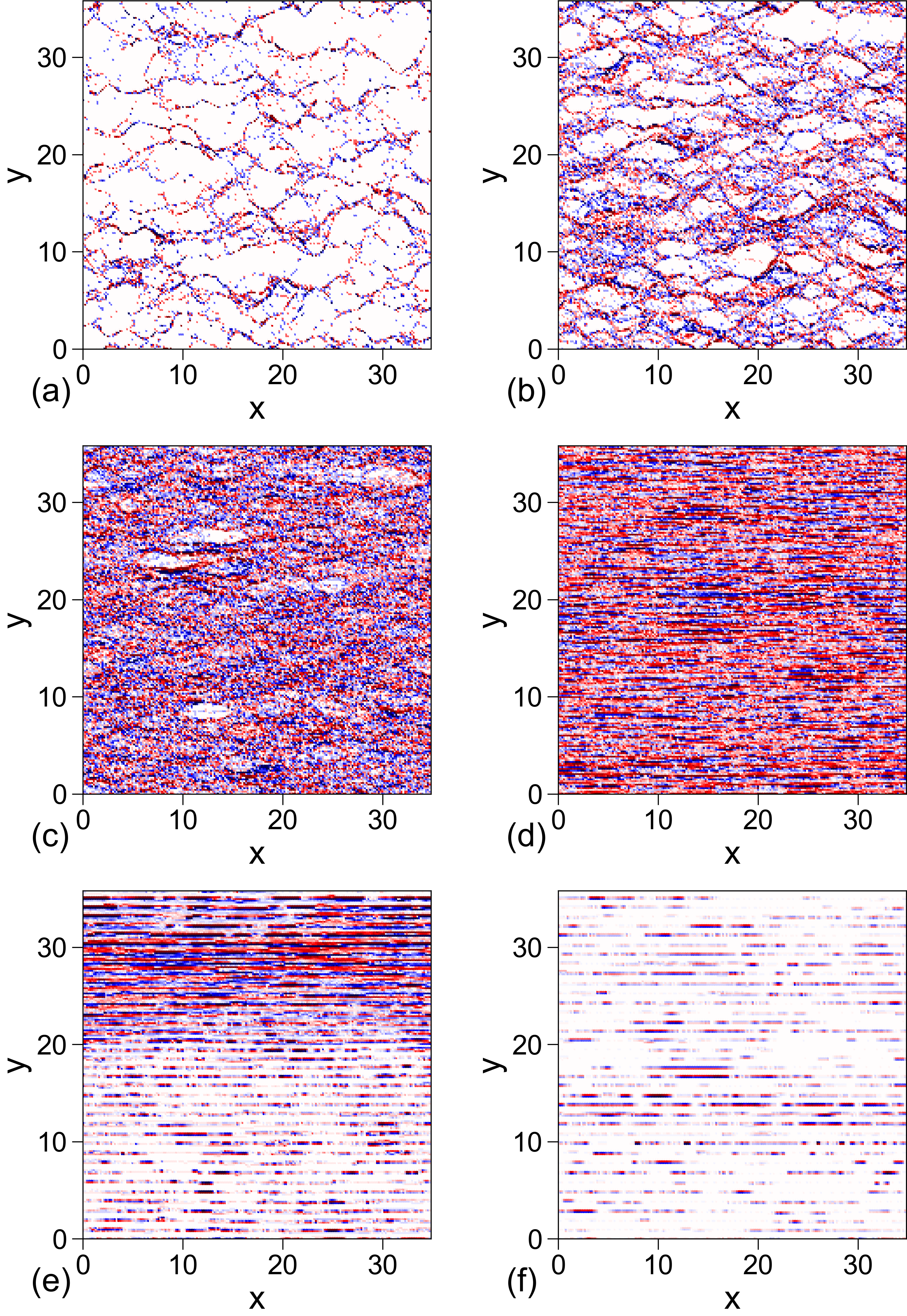}
\caption{Height map of the trajectory difference ${\bf D}_h$ as a
function of $x$ and $y$ for the $F_p=1.0$ system from Fig.~\ref{fig:4}.
White areas have a difference of zero, meaning that the flow is
either nonexistent in those regions or does not change over time.
Blue and red areas indicate a decrease or increase in the amount of
flow over time.
(a) $F_{D} = 0.23$ in the non-ergodic
plastic flow regime.
(b) At $F_{D} = 0.36$, the number of flow channels has increased
but there are still regions where flow is not occurring.
(c) $F_{D} = 0.5$, slightly below the transition from non-ergodic to ergodic
plastic flow.
(d) $F_{D} = 0.76$ in the ergodic plastic flow regime.
(e) $F_{D} =  1.0$, where the system has partially formed a
moving smectic. (f) $F_{D} = 2.0$ in the moving smectic state. 
}
\label{fig:5}
\end{figure}

In Fig.~\ref{fig:5}(a), we plot a height field of ${\bf D}_h$
for the $F_p=1.0$ and $F_D=0.23$ sample
from Fig.~\ref{fig:4}(b). There are large regions of white where no flow is
occurring, and the sparse channels are somewhat discontinuous since
there are bottleneck areas where the flow is nearly constant, giving
a trajectory difference of zero.
At $F_D=0.36$ in Fig.~\ref{fig:5}(b), numerous intersecting
channels are now present,
but there are still a large number of white regions where the flow is
either absent or unchanging,
so the plastic flow is non-ergodic.
In Fig.~\ref{fig:5}(c) at $F_{D}= 0.5$, slightly below the transition to
ergodic plastic flow, the motion is changing with time nearly everywhere
in the sample.
At $F_D=0.76$ in the ergodic plastic flow regime,
the system remains topologically disordered and Fig.~\ref{fig:5}(d) shows
that changes in the flow extend throughout the entire system.
At this value of $F_{D}$, the distribution of $v_x^i$ 
is bimodal, but individual vortices are no longer pinned even over
short time scales so there is no weight at $v_x^i=0$.
In Fig.~\ref{fig:5}(e) at $F_{D}= 1.0$, a moving smectic is beginning to
emerge, and a region of one-dimensional (1D) flow coexists with a region
that has more liquid-like flow that is two-dimensional in character.
Figure~\ref{fig:5}(f) shows the fully developed moving smectic
state at $F_{D} = 2.0$, where the vortices move in well-defined
1D channels.
The structure visible
in the moving smectic state
is due to the presence of dislocation pairs that permit
the individual channels of flow to slide past each another.

\begin{figure}
\includegraphics[width=\columnwidth]{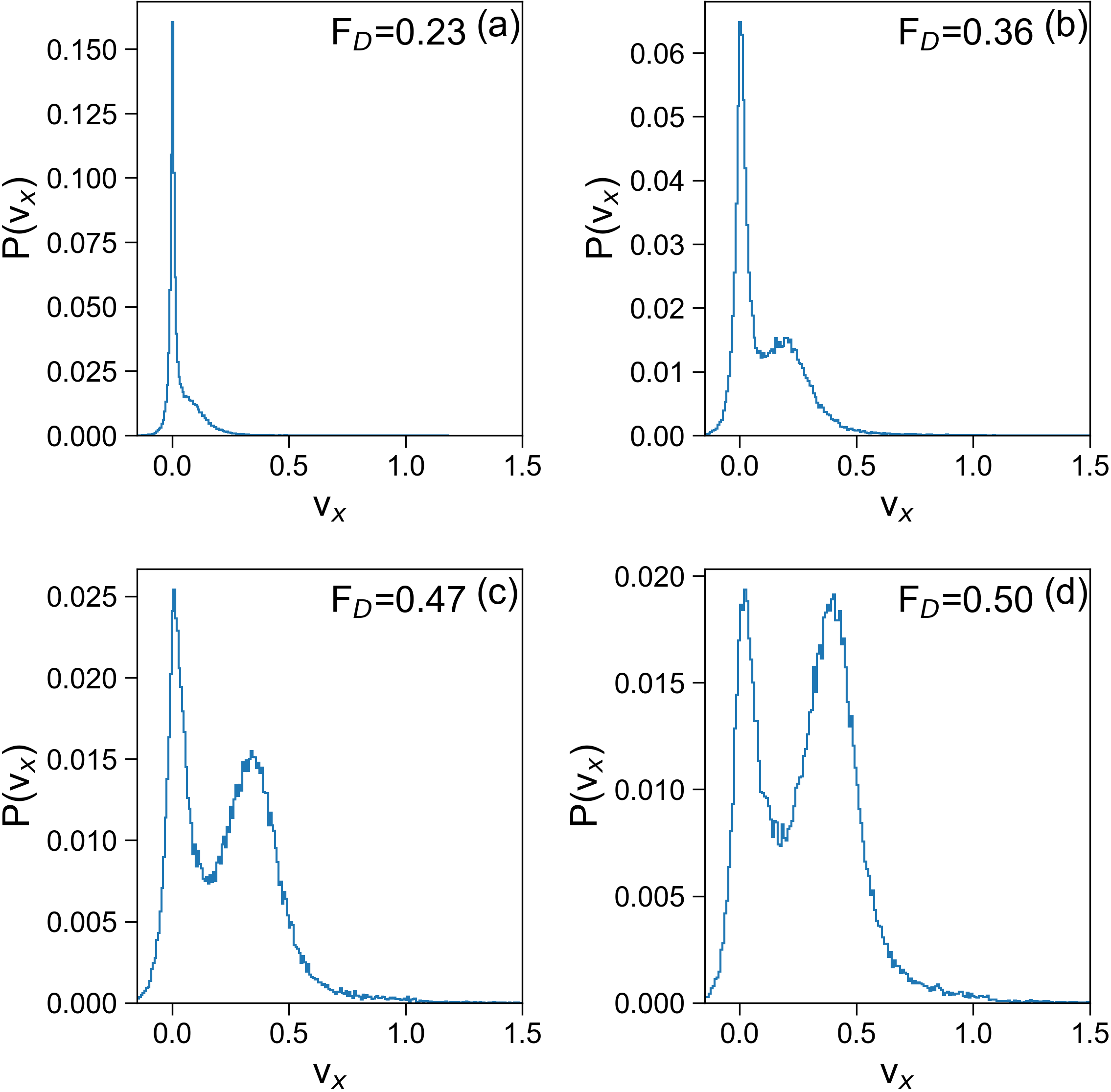}
\caption{Histograms of the distribution $P(v_x)$ of the instantaneous
vortex velocities $v_x$ for the $F_p=1.0$ system from Fig.~\ref{fig:1}
at drives in the non-ergodic plastic flow regime.
(a) $F_{D} = 0.23$.
(b) $F_{D} = 0.36$, where the velocity distribution has developed a
clear bimodal signature.
(c) $F_{D} = 0.47$, approaching the transition to ergodic plastic flow.
(d) $F_{D} = 0.5$, slightly below the ergodic plastic flow transition,
at the point where there is a peak in $P_2$ and a zero crossing in $P_3$.
The bimodal peaks in $P(v_x)$ have equal weight for this current.
}
\label{fig:6}
\end{figure}

To gain further insight into the different plastic flow phases,
we inspect the velocity distribution functions $P(v_x)$ for the instantaneous
$x$ velocities. A similar measurement appears in \cite{Faleski96}.
Figure~\ref{fig:6}(a) shows $P(v_x)$ for the $F_p=1.0$
system in Fig.~\ref{fig:5}
at $F_{D} = 0.23$ in the non-ergodic channel flow regime.
There is a large peak at $v_x=0$, indicating that
a large portion of the vortices remain permanently pinned,
along with
a finite positive $x$ velocity tail
representing the smaller number of vortices that
are able to move.
No second peak appears in $P(v_x)$,
indicating that the vortices traveling along the channels
do not have a characteristic velocity.
Instead, the motion through the channels is more pulse-like
or solitonic in character.
The shape of $P(v_x)$ remains similar
for lower values of $F_{D}$ such as $F_{D} = 0.125$.
At $F_{D} = 0.36$, where Fig.~\ref{fig:5}(b) illustrates that the channels are
less well defined and involve a greater amount of flow along the $y$
direction,
Fig.~\ref{fig:6}(b) shows that in addition to the large $v_x=0$ peak
in $P(v_x)$ from the permanently pinned vortices,
a second peak has emerged at finite velocity and the flow along the
channels has become continuous.
The second velocity
peak is centered near $v_{x} = 0.2$, which is smaller than the free
flow value of $v_x=F_D=0.36$ because 
the channels exhibit
strong meandering along the $y$ direction.
The interactions between the moving vortices and the vortices
that remain trapped in the pinning sites
produce an additional drag effect.
In Fig.~\ref{fig:6}(c) at $F_{D} = 0.47$,
there are two well-developed peaks in $P(v_{x})$.
The peak at $v_x=0$, originating from the permanently pinned
vortices, remains higher than the finite velocity peak from the
channel flow. This current is close to the point where $p_6$ reaches its
minimum value in Fig.~\ref{fig:1}(b).
At $F_D=0.5$, illustrated in
Fig.~\ref{fig:6}(d), the two peaks in $P(v_x)$ are very nearly of equal
height, but the maximum of the lower peak has just begun to shift above
$v_x=0$, indicating that the permanently pinned vortices are beginning
to disappear.
This is the same value of $F_D$ at which there is
a zero crossing of $P_3$ in Fig.~\ref{fig:1}(e), and it is slightly below
the drive at which $P_2$ peaks in Fig.~\ref{fig:1}(d).

\begin{figure}
\includegraphics[width=\columnwidth]{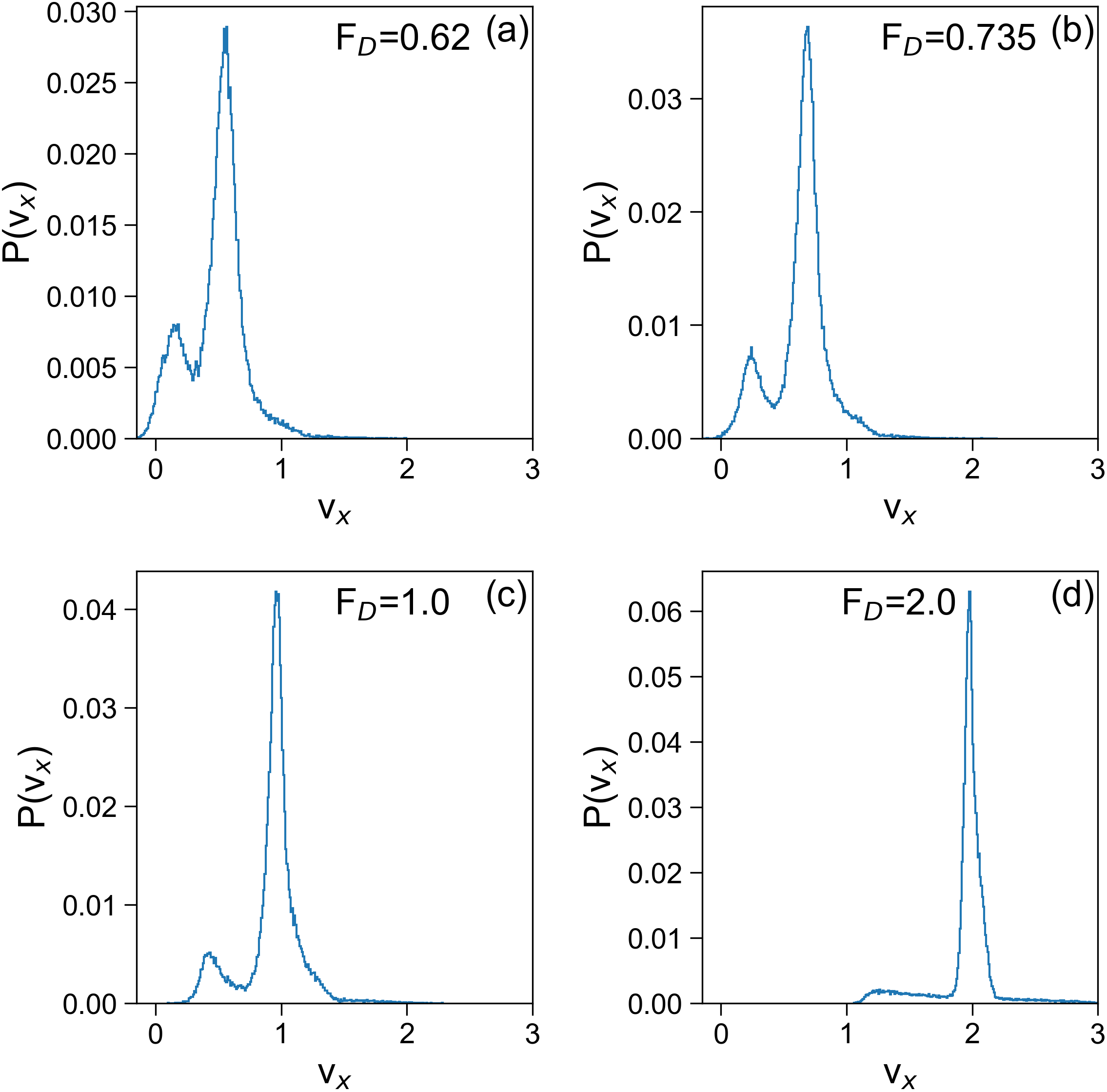}
\caption{Histograms $P(v_x)$ of the instantaneous vortex velocities
  $v_x$ for the $F_p=1.0$ system from Fig.~\ref{fig:1} at drives in the
  ergodic plastic flow regime and the dynamically ordered smectic state.
  (a) $F_{D} = 0.62$ at the transition from the non-ergodic to the
  ergodic plastic flow phase, where some vortices are temporarily pinned
  but no vortices are permanently pinned.
  (b) $F_{D} = 0.735$ in the ergodic plastic
  flow phase, where $P(v_x)$ remains bimodal but no vortices are
  pinned, so the weight near $v_x = 0$ is nearly zero.
  (c) $F_{D} = 1.0$ in the region illustrated in Fig.~\ref{fig:5}(e) where
  a dynamically reordered moving smectic coexists with a moving fluid.
  (d) $F_{D} = 2.0$ in the moving smectic state.
}
\label{fig:7}
\end{figure}

In Fig.~\ref{fig:7}(a),
we plot $P(v_{x})$ for $F_{D} = 0.62$ at the transition between the
non-ergodic and ergodic plastic flow states, corresponding to the drive at
which there is a zero crossing in $P_1$ and a local minimum in $P_3$ in
Fig.~\ref{fig:1}(c,e).
There are still some vortices that spend a short time being pinned,
as indicated by the finite weight at $v_x=0$,
but most of the vortices are moving at a velocity of $v_x = 0.55$.
The velocity distribution function remains strongly bimodal, so
the system is still topologically disordered and
individual vortices are able to exchange neighbors as they move.
At $F_D=0.735$ in Fig.~\ref{fig:7}(b),
the vortices are undergoing ergodic plastic flow
and $P(v_x)$ is bimodal; however, there is almost no weight at
$v_x = 0$,
indicating that although the pinning sites are able to slow the motion of
the vortices, they are no longer able to capture the vortices.
In Fig.~\ref{fig:7}(c) at $F_{D} = 1.0$, the drive overwhelms the pinning
force since $F_D=F_p$. The velocity distribution function is weakly
bimodal
with a small peak near $v_x = 0.43$
and another more prominent peak at $v_x=0.96$, just below $v_x  = F_D$.
As illustrated in Fig.~\ref{fig:5}(e), for this drive there is a coexistence
between a moving fluid state with a reduced average velocity corresponding
to the $v_x=0.43$ peak, and a moving smectic state, with an average velocity
corresponding to the $v_x=0.96$ peak.
At $F_D=2.0$, Fig.~\ref{fig:7}(d) shows $P(v_x)$ 
for the moving smectic phase, where there is a single
strong peak at $v_x=1.98$ just below $v_x = F_D$.
This peak sits on top of a low and very broad velocity distribution
produced by the dislocations in the smectic state, which gradually
translate backwards in the $-x$ direction through the assembly of vortices.
As individual vortices pass through the dislocations, they are slowed
in their motion.

Based on the velocity histograms, images, transport properties,
and signatures in the principal components $P_{n}$,
we suggest that the system passes through the following dynamic
phases with increasing drive:
(I) pinned, (II) plastic flow via pulse or soliton motion
through q1D channels surrounded by
a large number of permanently pinned vortices,
(III) continuous plastic flow
through quasistatic q1D channel phases
surrounded by some regions of permanently pinned vortices,
(IV) disordered
plastic flow where all vortices move throughout the system
but vortices can be pinned temporarily,
(V) a disordered moving fluid where the velocity distribution
remains bimodal but vortices do not spend any time being pinned,
(VI)
coexisting moving fluid and moving
smectic regions,
and (VII) a dynamically reordered moving smectic phase.
Non-ergodic plastic flow occurs in
phases II through IV, and there is ergodic plastic flow
in phase V.
The moving smectic state is also non-ergodic
since the vortices flow only along specific 1D paths.

\begin{figure}
\includegraphics[width=\columnwidth]{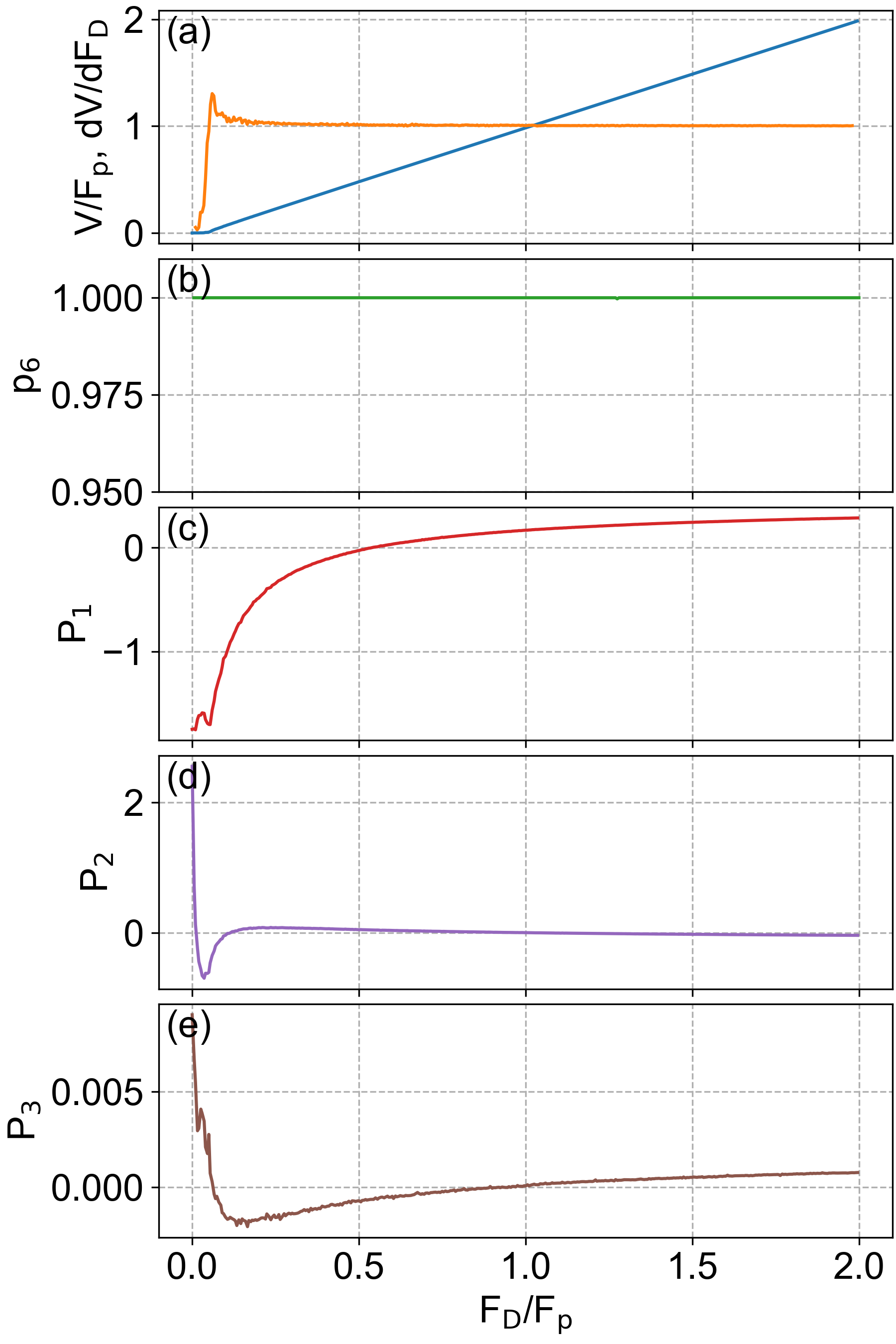}
\caption{(a) $V$ (blue) and $dV/F_{D}$ (orange) vs $F_{D}$ for
  an elastic depinning sample with $F_{p}= 0.05.$
  There is a pinned phase and a moving crystal state.
  (b) Corresponding $p_{6}$ vs $F_{D}$.
  For all values of $F_D$, $p_6=1.0$.
  (c-e) Corresponding PVB principal components $P_n$ vs $F_D$.
  (c) $P_1$. (d) $P_{2}$. (e) $P_{3}$.
}
\label{fig:8}
\end{figure}

We compare the plastic depinning results to two-dimensional
elastic depinning where the system forms a pinned
crystal that depins elastically without
generation of topological defects into a moving crystal state. 
In Fig.~\ref{fig:8}(a) we
plot $V$ and $dV/F_{D}$ versus $F_{D}$ for the same system
from Fig.~\ref{fig:1} at $F_{p}= 0.05$, in the elastic depinning
regime.
The plot of $P_6$ versus $F_D$ in Fig.~\ref{fig:8}(b)
shows that $P_6=1.0$ for all $F_{D}$,
indicating that there is no generation of defects.
There is a small peak in $dV/dF_{D}$ at the depinning transition
in Fig.~\ref{fig:8}(a),
and for higher drives
$dV/dF_D=1.0$.
In systems with elastic depinning, the velocity
scales with the driving force and depinning threshold $F_c$ as
$v = (F_{D} - F_{c})^\beta$ where $\beta < 1.0$,
with typical values of $\beta=0.66$, so that there is a nonlinear
velocity-force signature just above elastic depinning
\cite{Reichhardt17}.
Figure~\ref{fig:8}(c,d,e) shows the PVB PCA derived order parameters
$P_1$, $P_2$, and $P_3$ versus $F_{D}$.
In the elastic pinning regime, these order parameters contain
far fewer features than what we observe in Fig.~\ref{fig:1}(c,d,e)
for a system that undergoes plastic depinning.
The clearest transition identified by the $P_n$ is the depinning transition.
There is a zero crossing of $P_1$ at $F_D/F_p=0.5$, and
there are zero crossings
of $P_2$ and $P_3$ at $F_D/F_p=1.0$.
This suggests that the features in $P_{n}$ for plastic depinning are capturing
large scale changes in the structure and dynamics,
while the lack of features for elastic depinning reflects the fact
that there are no changes in
the lattice structure or in the dynamics above depinning.

\section{Varied Pinning Force}

\begin{figure}
\includegraphics[width=\columnwidth]{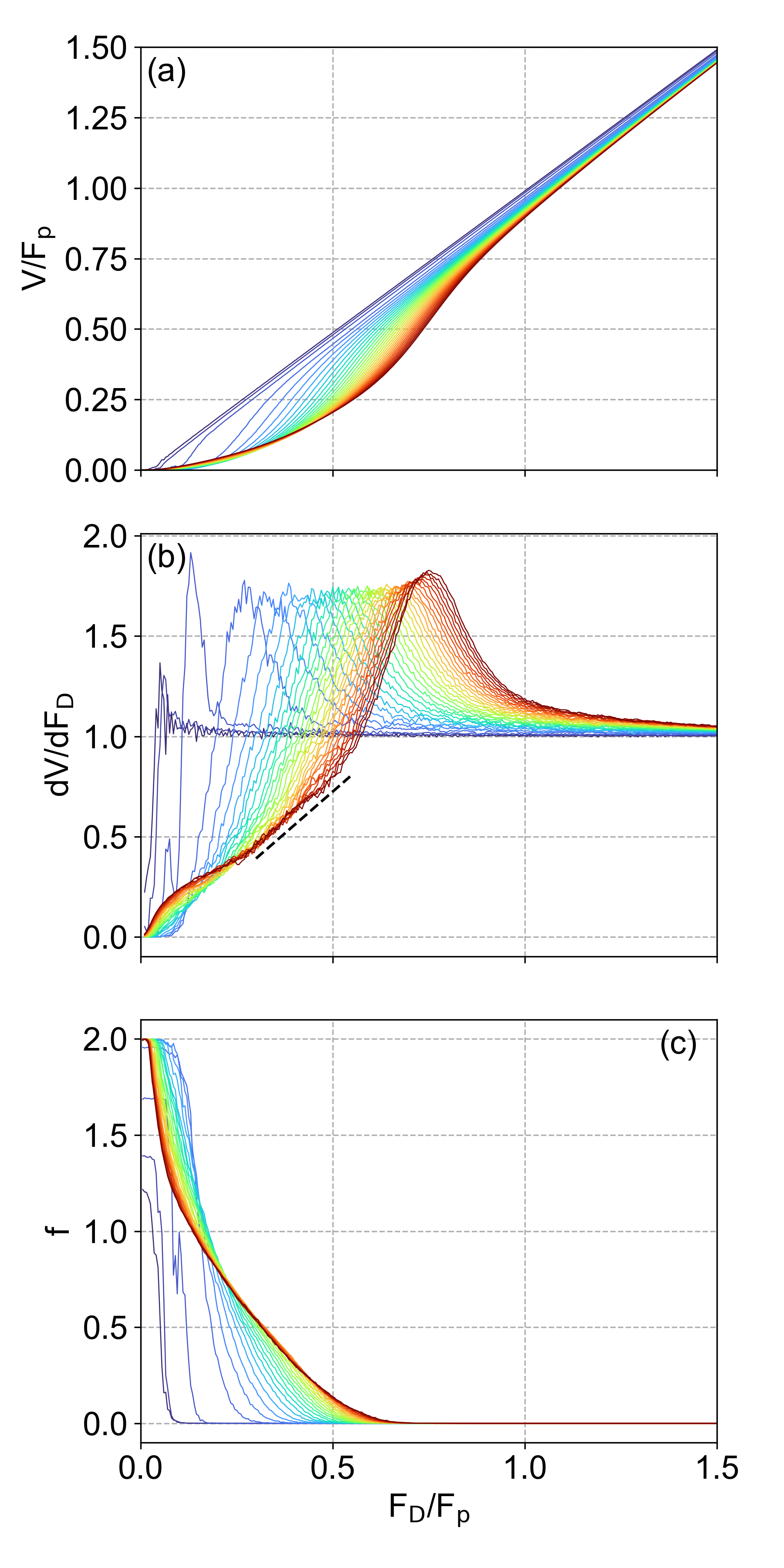}
\caption{(a) $V/F_{p}$ vs $F_{D}/F_{p}$
for the same system from Fig.~\ref{fig:1} at
$F_{p}= 0.025$ (dark blue, top), 0.05, 0.1, 0.2,
0.3, 0.4, 0.5, 0.6, 0.7, 0.8, 0.9, 1.0, 1.1, 1.2, 1.3, 1.4, 1.5, 1.6,
1.7, 1.8, 1.9, 2.0, 2.1, 2.2, 2.3, 2.4, 2.5, 2.6, 2.7, 2.8, 2.9,
and 3.0 (dark red, bottom).
(b) The corresponding $dV/dF_{D}$ vs $F_{D}/F_{p}$ curves.
The dashed black line is a linear fit,
offset slightly to the right for visibility,
to $dV/dF_D$
in the non-ergodic plastic flow regime for the larger $F_{p}$
values, indicating that $V \propto F_D^2$
in this plastic flow state.
(c) The corresponding pinned fraction $f$ vs $F_D/F_p$ curves.
}
\label{fig:9}
\end{figure}

We next consider how all of the measures
described above evolve as a function of $F_{p}$.
In Fig.~\ref{fig:9}(a)
we plot $V/F_{p}$ versus $F_{D}/F_{p}$ for the system from Fig.~\ref{fig:1}
over a range of
$F_{p}$ values from $F_p= 0.025$ to $F_p=3.0$.
Figure~\ref{fig:9}(b) shows the corresponding
$dV/dF_{D}$ versus $F_{D}/F_{p}$ curves.
We take $F_p$ values in increments of 0.1 from $F_p=0.1$ to $F_p=3.0$,
and also include $F_p=0.025$ and $F_p=0.05$ to access the elastic
depinning regime that appears for $F_p<0.1$.
For samples with $F_p \geq 1$ in the plastic depinning regime,
there is an extended region of $F_D/F_p$ values over which
$dV/dF_{D}$ increases linearly with $F_D/F_p$, and for higher drives,
$dV/dF_D$  increases more sharply
before reaching a peak and then gradually saturating towards
$dV/dF_D=1.0$.
In systems with plastic depinning,
$V \propto (F_{D} -F_{c})^\beta$ with $\beta > 1.0$,
where exponents of $1.5$ to $2.0$ are typically observed \cite{Reichhardt17}.
In Fig.~\ref{fig:9}(b), we
plot a linear fit to a portion of the curves with
large $F_p$ where
$dV/dF_D \propto F_D/F_p$, indicating that
$V \propto F_D^2$ and that therefore $\beta=2$.
There is a slope change in the $dV/dF_D$ curves for high $F_p$
where the curve ceases to follow the linear fit and begins to
increase more steeply with increasing $F_D/F_p$.
As shown by the $f$ versus $F_D/F_p$ curves plotted in
Fig.~\ref{fig:9}(c), this slope change occurs at the point where
$f=0$ for each of the $F_p$ values.
Together, these results suggest
that in the non-ergodic plastic flow regime,
where permanently pinned vortices are present,
the velocities scale as $V = F_D^2$,
while in the ergodic plastic flow regime,
vortices are at most temporarily pinned and the
velocity-force curves fall into a different scaling regime.

\begin{figure}
\includegraphics[width=\columnwidth]{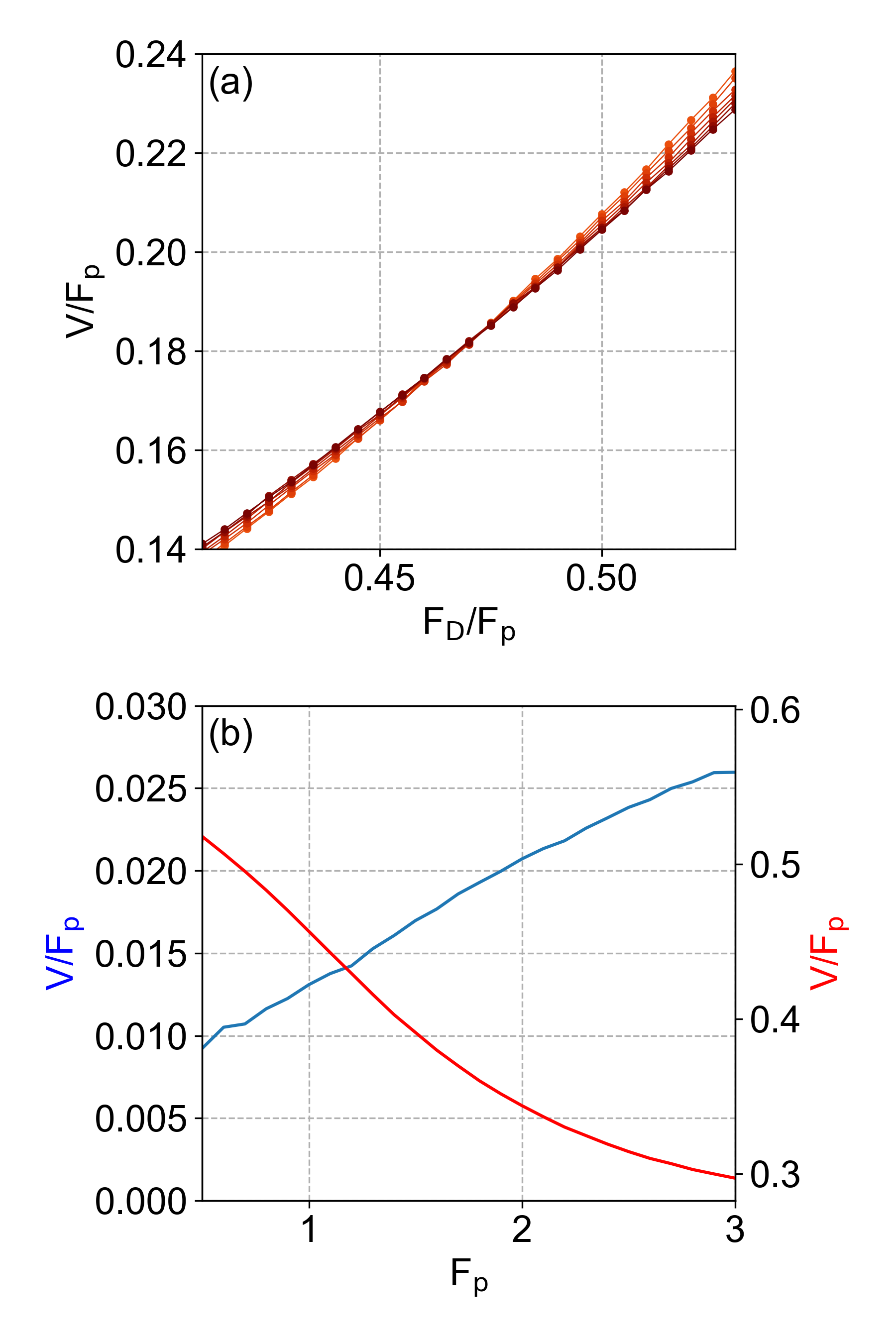}
\caption{(a) Zoomed in plot of $V/F_p$ vs $F_D/F_p$ data from
  Fig.~\ref{fig:9}(a) for $F_p \geq 2.4$ showing a crossing of
  the scaled curves.
(b) $V/F_p$ vs $F_{p}$ at fixed drives of
  $F_{D}/F_{p} = 0.15$ (blue) where the velocity increases with
  increasing $F_p$ and
  $F_{D}/F_{p} = 0.6$ (red) where the  velocity decreases with
  increasing $F_p$.
}
\label{fig:10}
\end{figure}

In Fig.~\ref{fig:9}(a), we find that there is a crossing of the scaled
$V/F_p$ versus $F_D/F_p$ curves when $F_p \geq 2.4$, meaning that
for a fixed value of $F_{D}/F_{p}$,
the relative velocity counterintuitively increases
with increasing $F_{p}$.
The curves cross at
$F_{D}/F_{p} = 0.475$, as shown in Fig.~\ref{fig:10}(a)
where we plot a blow up of the $F_p \geq 2.4$ curves from
Fig.~\ref{fig:9}(a) around the crossing region.
Although the curves for lower $F_p$ do not cross at a single
value of $F_D/F_p$, they still cross, as we illustrate in
Fig.~\ref{fig:10}(b) where we plot $V/F_p$ versus $F_p$ for
fixed drive values of
$F_{D}/F_{p} = 0.15$ and $F_D/F_p=0.6$.
For the low drive of $F_D/F_p=0.15$, the
scaled velocity
increases from $V/F_p=0.009$ at $F_{p} = 0.5$
to $V/F_p=0.026$ at $F_{p} = 3.0$.
At the higher drive of
$F_{D}/F_{p} = 0.6$, the scaled velocity drops
from $V/F_p=0.52$ at $F_p=0.5$ to $V/F_p=0.3$ at $F_p=3.0$.

\begin{figure}
\includegraphics[width=\columnwidth]{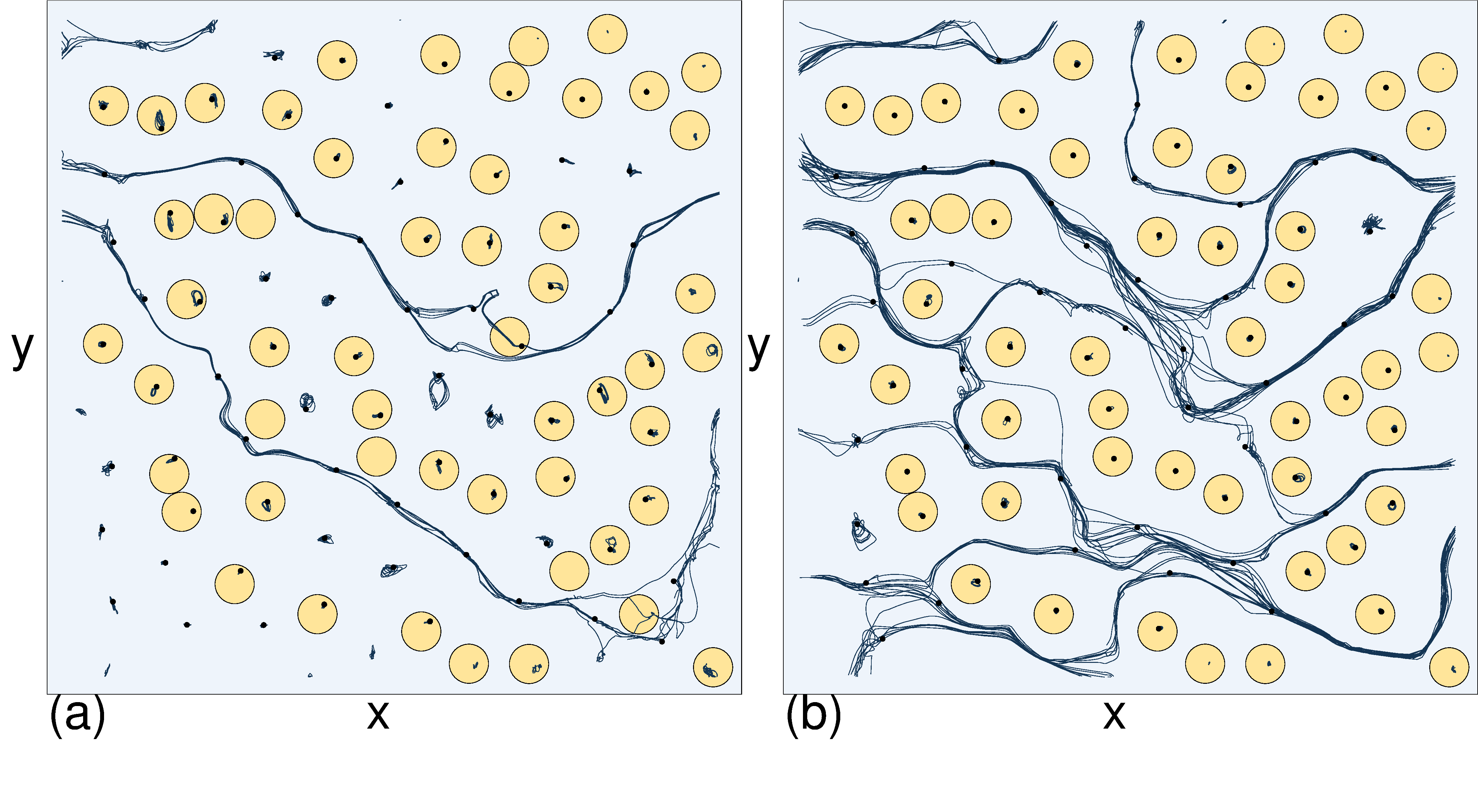}
\caption{Images of pinning sites (open circles), vortex positions (dots),
  and vortex trajectories (lines) over a time interval that has been
  scaled by $F_p$ at $F_D/F_p=0.125$
  in a small region of the system from Fig.~\ref{fig:10}.
  (a) At $F_{p} = 0.4$, the vortices flow at low velocity through
  channels. 
  There is also circular motion of the vortices
  that are trapped in the pinning sites and the interstitial regions.
  (b) At $F_{p} = 3.0$, the vortices flow rapidly through the channels,
  and the trajectories are imaged
  over a time interval that is 7.5 times shorter
  than the interval shown in panel (a).
  The circular motion of the vortices trapped in pinning sites
  and interstitial regions is significantly reduced, so the vortices
  in the flowing channels experience a reduced amount of drag compared
  to the system with lower $F_p$.
}
\label{fig:11}
\end{figure}

In Fig.~\ref{fig:11}(a), we show the pinning site positions, vortex positions,
and vortex trajectories in a small region of a sample with $F_p=0.4$ at
$F_{D}/F_{p} = 0.125$ in the non-ergodic plastic flow regime.
The vortices are traveling along a limited number of narrow 1D channels.
Adjacent vortices trapped in pinning sites undergo circular motion, and there
are also numerous pinned interstitial vortices that also undergo circular
motion as the vortices in the plastic channels flow past.
The ability of these two types of trapped vortices to move inside the
regions in which they are confined saps energy away from the vortices flowing
in the channels and slows their motion.
At $F_p=3.0$ and $F_D/F_p=0.125$,
Fig.~\ref{fig:11}(b) shows that the vortices flow through a greater number
of relatively wide 1D channels. There are almost no trapped interstitial
vortices because the interstitial trapping force is determined by the
strength of the vortex-vortex interaction, which remains constant because
we are holding the vortex density fixed. There are still permanently pinned
vortices present, but these vortices are heavily constrained by the strong
pinning sites and can only make small motions in response to the vortices
flowing past through the plastic channels. As a result, the pinned vortices
are only weakly coupled to the flowing vortices and slow the
motion of the plastically flowing vortices by a small amount.
This explains why $V/F_p$ is an increasing function of $F_p$ for
the small drive regime represented by
$F_D/F_p=0.15$ in Fig.~\ref{fig:10}(b).
In the large drive regime illustrated by $F_D/F_p=0.6$ in
Fig.~\ref{fig:10}(b), all of the vortices are flowing and there are no
vortices permanently trapped in pinning sites or interstitial sites.
Instead the vortices flow across the pinning sites. As a result,
increasing $F_p$ increases the coupling between the pinning sites and
the moving vortex assembly, increasing the drag on the assembly and
decreasing $V/F_p$.

\begin{figure}
\includegraphics[width=\columnwidth]{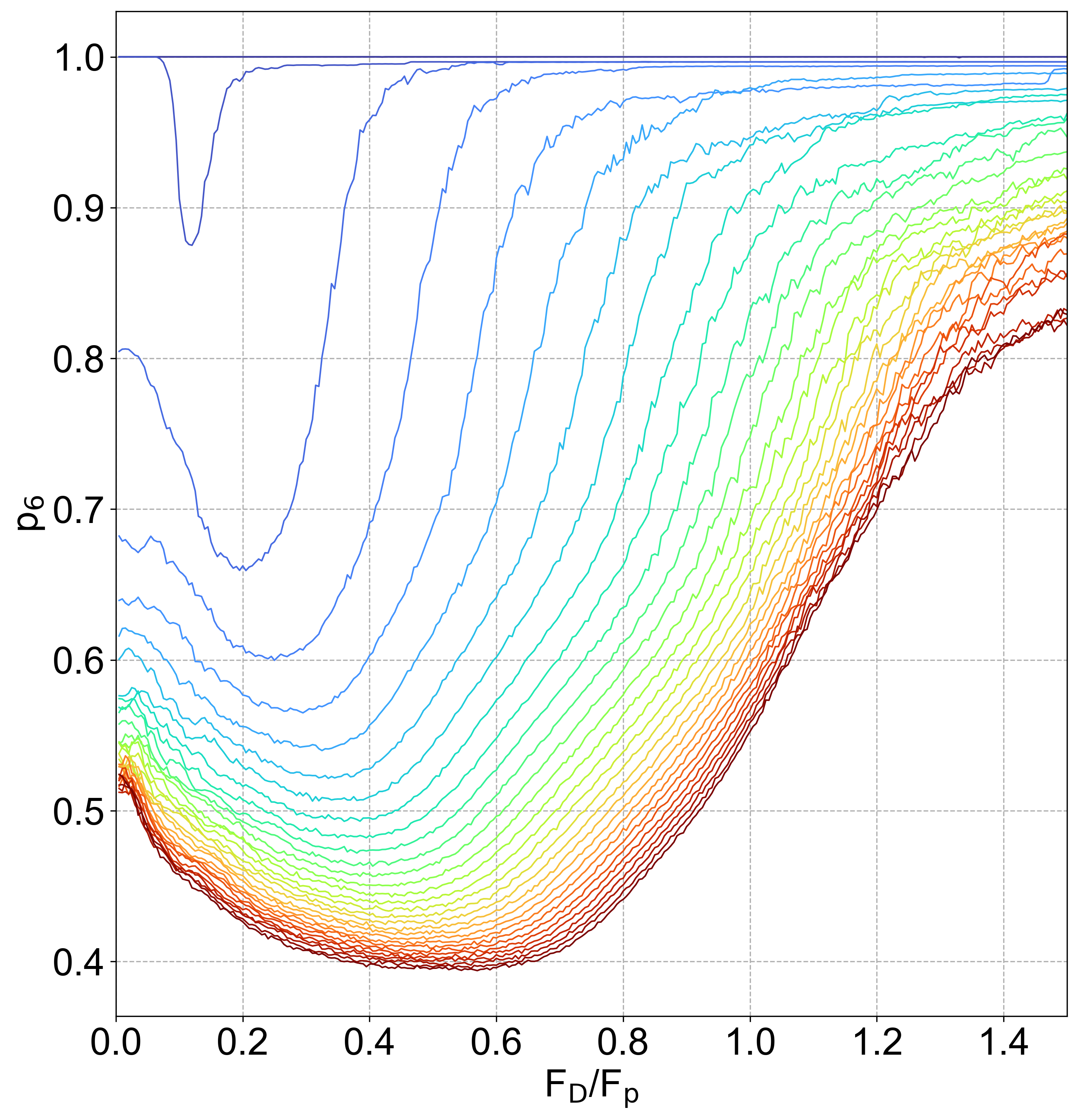}
\caption{The fraction of sixfold coordinated vortices $p_{6}$ vs $F_{D}/F_{p}$ at
$F_{p}= 0.025$ (dark blue, top), 0.05, 0.1, 0.2,
0.3, 0.4, 0.5, 0.6, 0.7, 0.8, 0.9, 1.0, 1.1, 1.2, 1.3, 1.4, 1.5, 1.6,
1.7, 1.8, 1.9, 2.0, 2.1, 2.2, 2.3, 2.4, 2.5, 2.6, 2.7, 2.8, 2.9,
and 3.0 (dark red, bottom).
For $F_{p} < 0.6$, the system dynamically orders into a moving crystal.
}
\label{fig:12}
\end{figure}

\begin{figure}
\includegraphics[width=\columnwidth]{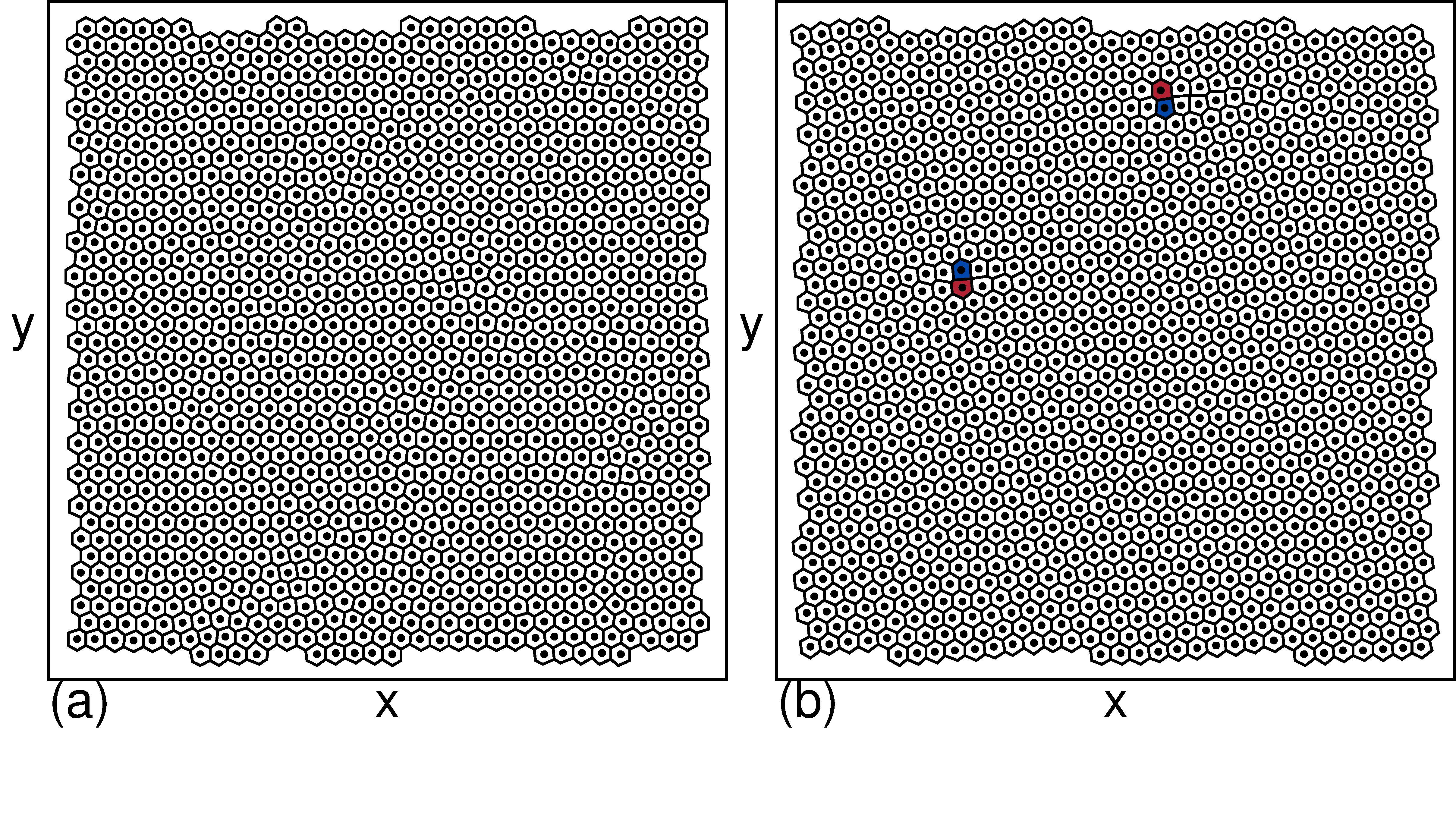}
\caption{The Voronoi construction for the system
in Fig.~\ref{fig:12} with $F_p=0.1$. (a)
$F_{D}/F_p= 0.02$.
(b)
$F_{D}/F_{p} = 0.4$.
In this case, the system forms a moving crystal rather than a moving smectic.
}
\label{fig:13}
\end{figure}

In Fig.~\ref{fig:12} we plot $p_{6}$ versus $F_D/F_{p}$ for varied $F_P$.
For the smallest $F_{p}$, the
system behaves elastically, and as $F_{p}$ increases,
an increasing number of defects appear
in the plastic flow phase.
For $F_{p} = 0.05$, the pinned state contains no defects,
giving $p_6=1.0$.
There is a small window of driving force over which about ten percent of
the lattice becomes defected, but $p_6$ quickly increases back to
$p_6 \approx 1.0$ for higher drives.
For $F_{p} = 0.1$, there is
a finite number of defects
present in the lattice for $F_{D}/F_{p} = 0$.
As $F_D$ increases,
$p_6$ passes through a minimum value in the plastic flow regime,
and then increases again
in the dynamical reordering regime.
For $F_{p} < 0.6$, the dynamically reordered state
is not a moving smectic but
is instead a moving crystal containing
only a few topological defects.
In Fig.~\ref{fig:13}(a) we show
a Voronoi construction of the defect-free moving state at
$F_{p} = 0.1$ and $F_{D}/F_p = 0.02$, while
in Fig.~\ref{fig:12}(b) we illustrate the same system
at $F_{D}/F_p = 0.04$, where almost no defects are present.
In general, the moving crystal or moving floating crystal
states can be distinguished from the moving smectic
state because they either contain
no defects or they have only a vanishingly small number of defects that
do not have their Burgers vector aligned exactly parallel with the
driving force.
For samples with stronger pinning of $F_p > 0.6$,
the high drive state is a
moving smectic containing aligned defects that increase in density with
increasing $F_p$.

The moving crystal state is better described as a floating
solid. In systems with quenched disorder,
it is known that when the coupling to the substrate is weak,
an Aubry transition can occur, and
it was shown that for 1D incommensurate chains on a substrate,
there is a second-order pinned to unpinned
transition at a critical disorder strength \cite{Peyrard83}.
The Aubry transition has now been observed in 2D systems for
colloidal particles on periodic lattices \cite{Brazda18},
cold atoms \cite{Bylinskii16}, and friction \cite{Kiethe17}.
An open question is whether an Aubry transition also occurs for a system
containing random disorder.
Our results suggest that
for sufficiently low pinning strengths,
the moving vortex lattice
does not form a moving smectic state
as the Burgers vectors of the defect pairs
are not aligned with the driving direction.
In this case, the lattice
is tilted slightly away from the direction of drive,
as shown in Fig.~\ref{fig:13}(b), because the elastic vortex-vortex
interactions dominate over the interactions with the substrate.
In the low $F_p$ range of the moving smectic phase,
there could be
a dynamical Aubry transition at higher drives consisting of a
second-order transition from a moving smectic to a moving crystal.
In other words, an Aubry transition might occur as a function of
driving force rather than pinning force.

\begin{figure}
\includegraphics[width=\columnwidth]{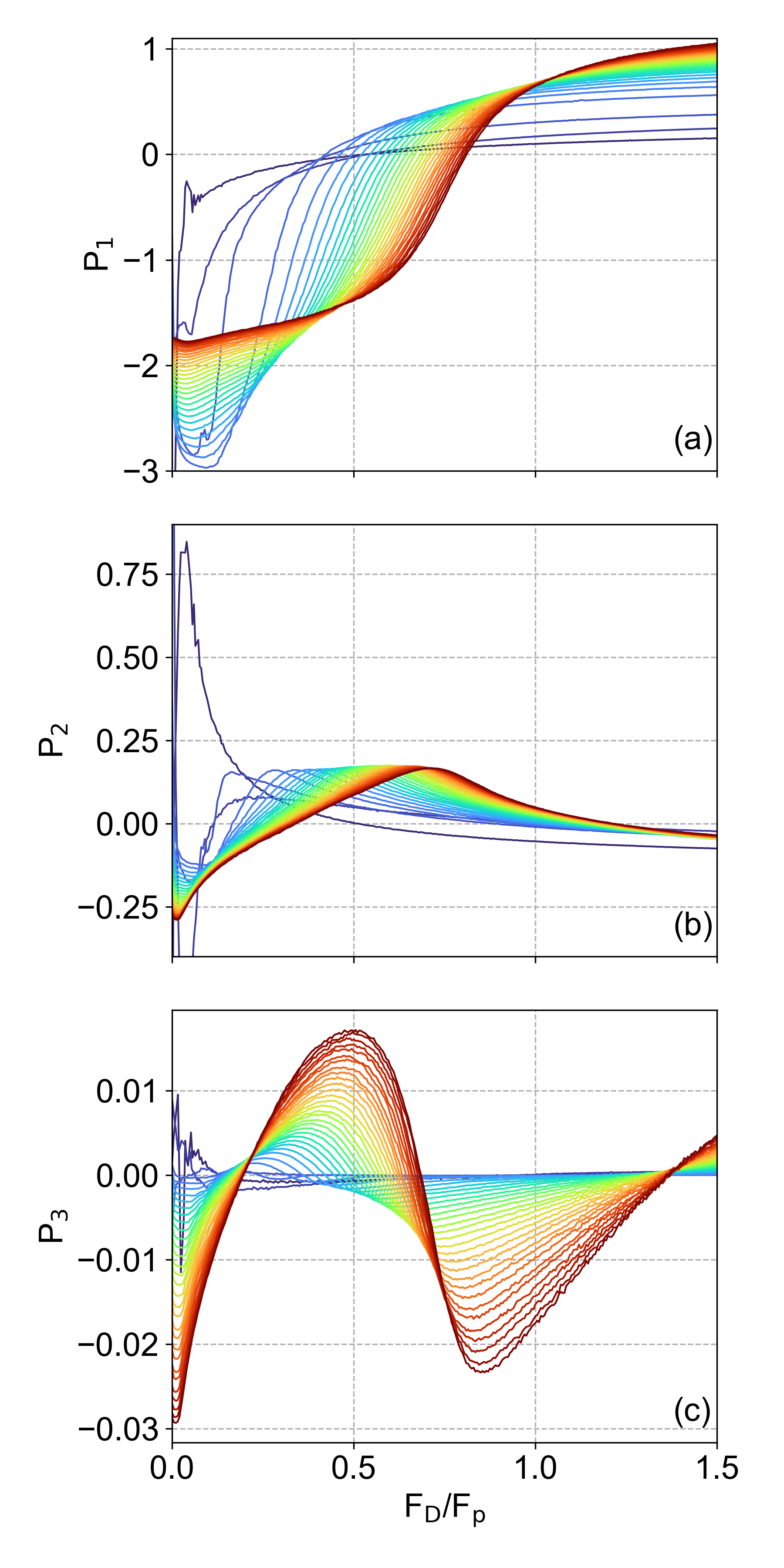}
\caption{PVB principal components for 
$F_{p}= 0.025$ (dark blue, bottom right), 0.05, 0.1, 0.2,
0.3, 0.4, 0.5, 0.6, 0.7, 0.8, 0.9, 1.0, 1.1, 1.2, 1.3, 1.4, 1.5, 1.6,
1.7, 1.8, 1.9, 2.0, 2.1, 2.2, 2.3, 2.4, 2.5, 2.6, 2.7, 2.8, 2.9,
and 3.0 (dark red, top right).
(a) $P_{1}$ vs $F_{D}/F_{p}$.
(b) $P_{2}$ vs $F_{D}/F_{p}$.
(c) $P_{3}$ vs $F_{D}/F_{p}$.
}
\label{fig:14}
\end{figure}

In Fig.~\ref{fig:14}(a), we plot the first
PVB principal component
$P_1$ versus $F_{D}/F_{p}$ for the system
from Fig.~\ref{fig:9} at varied $F_{p}$.
For $F_p=0.025$ and $F_p=0.05$, the depinning is elastic, and we find
that $P_1$ exhibits a peak at the depinning transition, a zero
crossing at $F_D/F_p=0.5$, and has
no other features for higher drives.
For larger $F_{p}$,
there is an initial dip in $P_1$ that
is associated with the plastic depinning transition,
followed by an extended
negative region in the non-ergodic plastic flow phase.
At a drive that increases with increasing $F_p$
and that falls in the range $0.2 < F_D/F_p < 0.6$, the upward slope of $P_1$
becomes significantly steeper and $P_1$ passes through a zero crossing
followed by a plateau at the higher drives.
The broad shapes of the $P_1$ curves
generally resemble that of the $p_6$ versus
$F_D/F_p$ curves in Fig.~\ref{fig:12}, but the details of the
$P_1$ and $p_6$ curves differ.
For large $F_{p}$, the non-ergodic plastic flow regime corresponds to the
lower slope and low drive portion of $P_1$,
the ergodic plastic flow regime occurs in the region of large slope,
and the dynamic ordering occurs on the high drive plateau region.
In Fig.~\ref{fig:14}(b) we plot the corresponding
second PVB principal component $P_{2}$ versus $F_{D}/F_{p}$,
where in the plastic flow regime we find
a local peak that shifts to higher $F_{D}/F_{p}$
with increasing $F_{p}$.
The peak is correlated with the crossover at which permanently pinned
vortices cease to be present, which also corresponds to the
transition from non-ergodic to ergodic plastic flow.
At higher drives, the zero crossing of $P_2$ occurs at the point where
the system begins to form a moving smectic state.

In Fig.~\ref{fig:14}(c)
we plot the third PVB principal component
$P_{3}$ versus $F_{D}/F_{p}$
for varied $F_{p}$.
For large $F_p$, there is a clear dip at the depinning threshold,
a local maximum that approaches $F_{D}/F_{p} = 0.5$,
and a local minimum that approaches
$F_{D}/F_{p} = 0.85$.
In the plastic depinning regime, there are three zero crossings
in the range of $0.12 < F_D/F_p<0.19$,
$0.32 < F_D/F_p<0.68$,
and $1.1 < F_D/F_p < 1.36$.
These features suggest that at least for large $F_{p}$,
$P_{3}$ is able to capture the
crossovers between the different phases.

\begin{figure}
\includegraphics[width=\columnwidth]{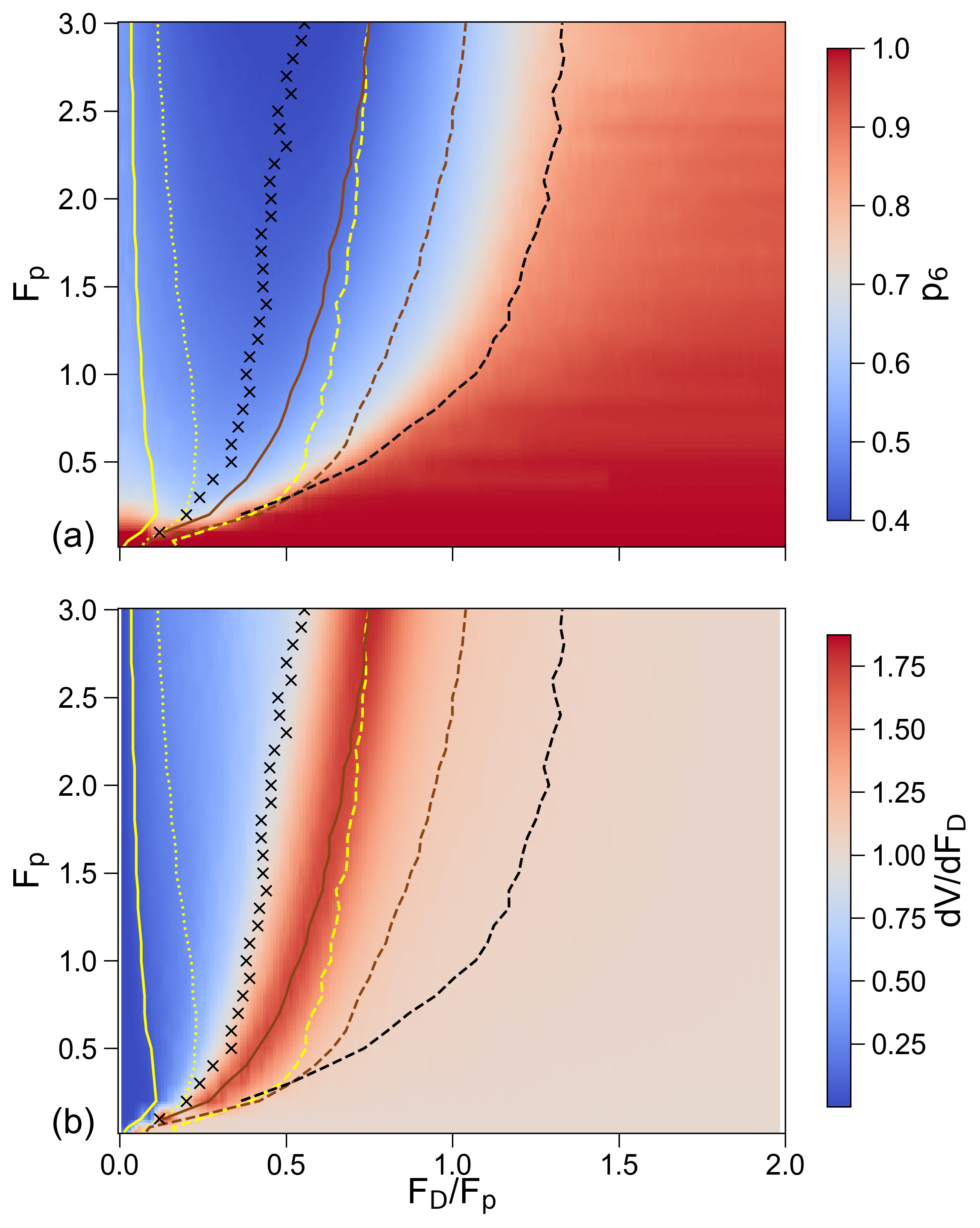}
\caption{(a) Height map of $p_6$ as a function of
$F_{p}$ vs $F_{D}/F_{p}$.
(b) Height map of $dV/dF_D$ as a function of
$F_p$ vs $F_D/F_p$.
Marked on both plots are: 
depinning threshold (solid yellow line),
$f_{ip}=0$ or the loss of
permanently pinned interstitial particles (dotted yellow line),
minimum in $p_6$ (black $x$'s),
peak in $dV/dF_D$ (solid brown line),
$f=0$ or the loss of
all permanently pinned particles (dashed yellow line),
flattening of $dV/dF_D$
(dashed brown line),
$p_6$ reaches 90\% of its maximum value (dashed black line).}
\label{fig:15}
\end{figure}

We next construct heat maps
of the different quantities to
get a better picture of the correlations among them.
In Fig.~\ref{fig:15}(a)
we show the height map of $p_6$ as a function of
$F_{p}$ vs $F_{D}/F_{p}$.
Marked on the figure are the depinning curve, the point at which there are
no longer any vortices permanently pinned
in interstitial locations, the point where
$p_6$ passes through its minimum value, the location of the
peak in $dV/dF_{D}$, the point at which there are no longer any
vortices permanently pinned in pinning sites,
the point where $dV/dF_{D}$ flattens, and
the point where $p_6$ reaches 90\% of its maximum value.
The system is ordered and depins elastically into a moving crystal state
for $F_{p} < 0.1$.
For high $F_p$, the peak in $dV/dF_D$ begins to overlap with
the curve indicating that
there are no longer vortices permanently pinned in pinning sites.
The minimum in $p_{6}$ falls well below the point at which there are
no longer permanently pinned vortices.
We also find that the depinning threshold drops to lower values of $F_D/F_p$
with increasing $F_p$ for high values of $F_p$ due to the reduced drag
on the flowing vortices by the highly confined pinned vortices,
illustrated in 
Fig.~\ref{fig:11}.
Figure~\ref{fig:15}(b) shows a heat map of $dV/F_{D}$ as a
function of $F_{p}$ versus $F_{D}/F_{p}$.
The minimum value of $p_6$ runs roughly parallel to the line
along which $dV/dF_D=1$.
The line marking the flattening of $dV/dF_{D}$ is associated with the
emergence of a moving smectic state,
but this is a continuous transition
and not a sharp transition, so the line is not sharply defined and
does not coincide with any signatures in $p_6$ or $dV/dF_D$.

\begin{figure}
\includegraphics[width=0.9\columnwidth]{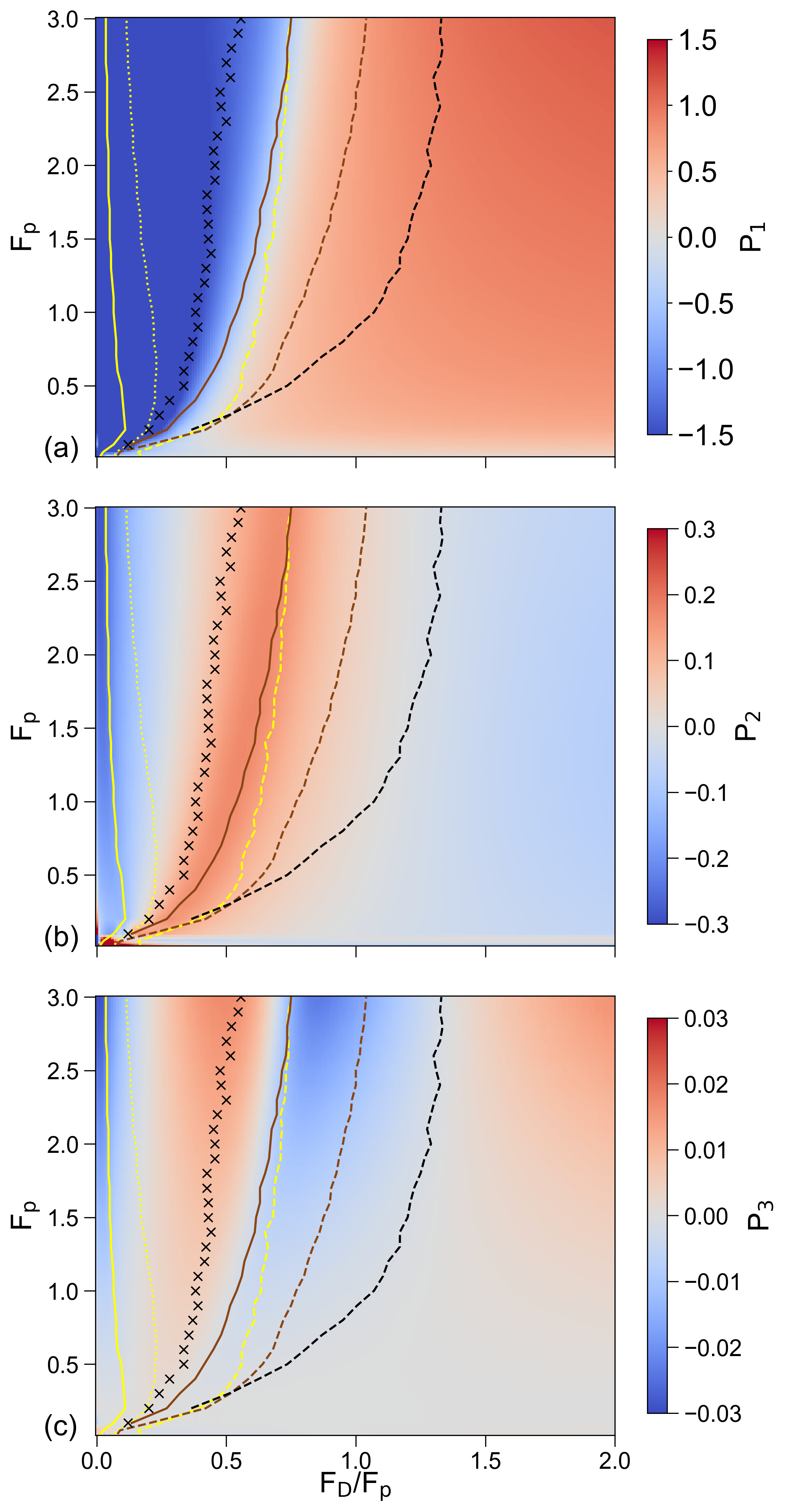}
\caption{(a) Height map of $P_1$ as a function of
$F_{p}$ vs $F_{D}/F_{p}$.
(b) Height map of $P_2$ as a function of
$F_p$ vs $F_D/F_p$.
(c) Height map of $P_3$ as a function of
$F_p$ vs $F_D/F_p$.
Marked on all plots are: 
depinning threshold (solid yellow line),
$f_{ip}=0$ or the loss of
permanently pinned interstitial particles (dotted yellow line),
minimum in $p_6$ (black $x$'s),
peak in $dV/dF_D$ (solid brown line),
$f=0$ or the loss of
all permanently pinned particles (dashed yellow line),
flattening of $dV/dF_D$
(dashed brown line),
$p_6$ reaches 90\% of its maximum value (dashed black line).}
\label{fig:16}
\end{figure}

In Fig.~\ref{fig:16}(a) we plot
a heat map of the PVB $P_1$
as a function of $F_{p}$ versus $F_{D}/F_{p}$.
The point at which $P_1$ begins to increase more rapidly from its
negative value towards zero coincides with the minimum in $p_6$.
The zero crossing line of $P_1$ falls between
the peak in $dV/dF_{D}$
and the line indicating that there are
no longer any permanently pinned vortices, two measures that
merge at high $F_p$.
All of these signatures are lost for very low $F_p$ where the flow is
elastic for all drives above depinning.
Figure~\ref{fig:16}(b) shows the heat map of the PVB $P_2$
as a function of $F_p$ versus $F_D/F_p$.
The minimum $P_2$ at small $F_D/F_p$ coincides with the
depinning threshold.
The zero crossing line of $P_2$ falls slightly below the minimum in
$p_6$, while the maximum in $P_2$ is near but does not follow
the peak in $dV/dF_D$, as we discuss in more detail below.
At high drives, the
maximum in $P_2$ approaches the
point at which there are no longer any permanently
pinned vortices.
A second zero crossing of $P_2$
occurs near the line along which $dV/dF_D$ begins
to flatten and saturate.

In Fig.~\ref{fig:16}(c), we show the height map of
the PVB $P_3$ as a function of $F_{p}$ versus
$F_{D}/F_{p}$.
The minimum in $P_3$ at low $F_D$ becomes deeper as $F_p$ increases, and
there is a clear signature of the depinning transition. The line marking
the point at which there are no longer any permanently pinned interstitial
vortices is near the first zero crossing of $P_3$. The nonergodic plastic
flow regime occurs in the window where $P_3$ is positive, and ends at the
second zero crossing of $P_3$. This crossing falls between the minimum
in $p_6$ and the peak in $dV/dF_D$, and has not been detected with any
previously considered measure to our knowledge.
In the window where the value of $P_3$ is negative, we find ergodic
plastic flow, and this flow ends at the crossover to the moving
smectic state, marked by the flattening of $dV/dF_D$ and by another zero
crossing region of $P_3$. This zero crossing occurs over a quite wide
region, consistent with the idea that the transition back into nonergodic
flow at high drives is a gradual crossover. For large $F_D$ and large $F_p$,
$P_3$ becomes positive again in the nonergodic moving smectic flow state.

The features from the height maps indicate that
our PVB based PCA technique can detect
signatures of the different dynamic transitions at least as well
as standard measures and, in some cases, produces
even more pronounced features at the transitions.
We have also examined height maps of $P_1$ and $P_2$ constructed using
purely particle based (PB) PCA. The data is much noisier and only captures
the dynamic reordering of the lattice, without finding any details inside
the plastic flow state.

\begin{figure}
\includegraphics[width=\columnwidth]{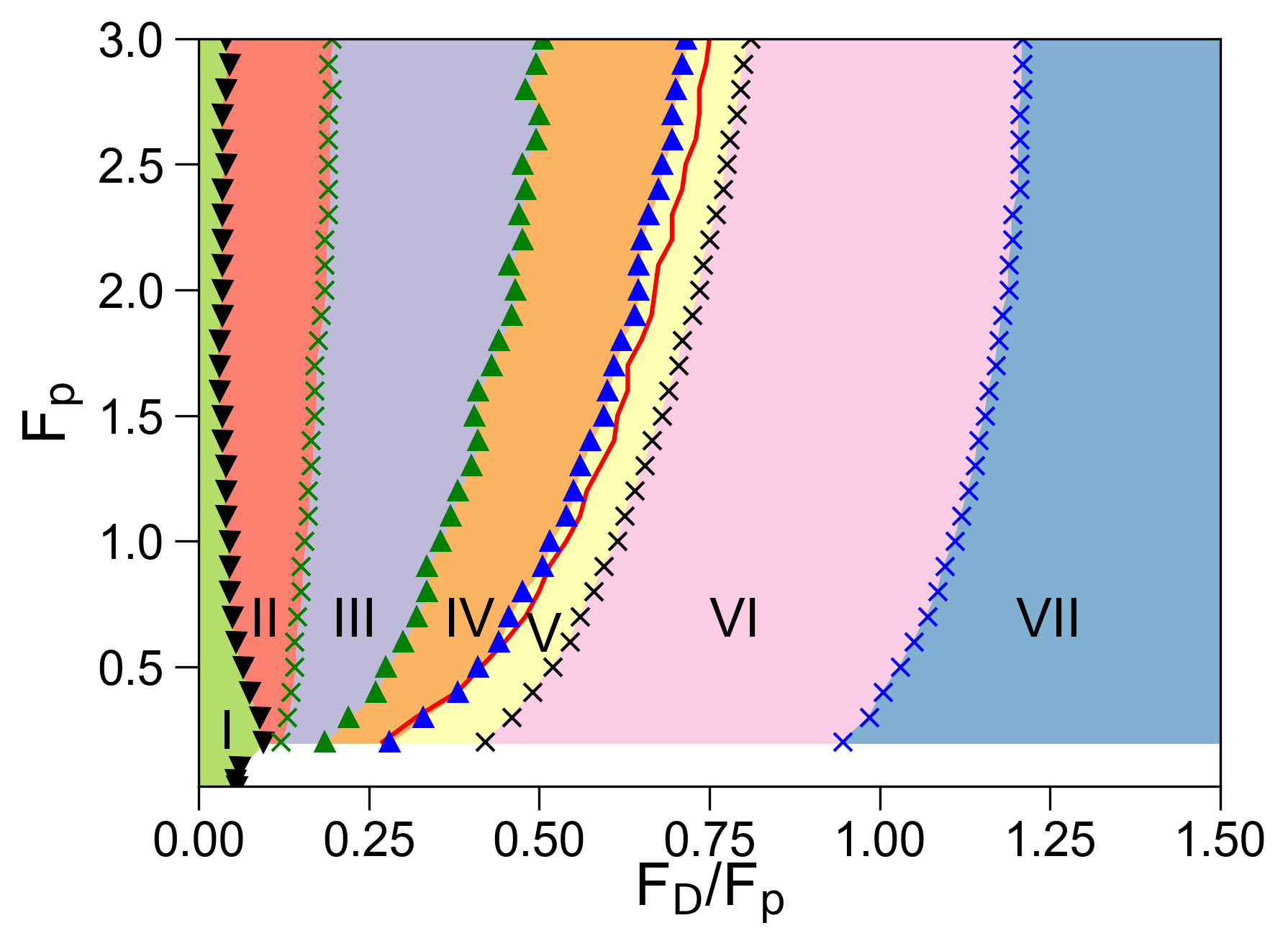}
\caption{Phase diagram as a function of $F_p$ vs $F_D/F_p$ constructed using
the features identified by PVB PCA.
Green: Pinned (phase I).
Red: Isolated channel flow (phase II).
Purple: Lightly braided channel flow (phase III).
Orange: Heavily braided channel flow (phase IV).
Yellow: Inhomogeneous ergodic plastic flow (phase V).
Pink: Emerging one-dimensional flow (phase VI).
Blue: Dynamically reordered smectic state (phase VII).
The phase boundaries are obtained from the following measures:
I-II (black down triangles): minimum of $P_1$.
II-III (green crosses): lowest zero crossing of $P_3$.
III-IV (green up triangles): peak of $P_3$.
IV-V (blue up triangles): peak of $P_2$.
V-VI (black crosses): zero crossing of $P_1$.
VI-VII (blue crosses): upper zero crossing of $P_3$.
The solid red line indicates the peak in $dV/dI$, which does not
directly coincide with the IV-V phase boundary.}
\label{fig:17}
\end{figure}

\begin{figure}
  \includegraphics[width=\columnwidth]{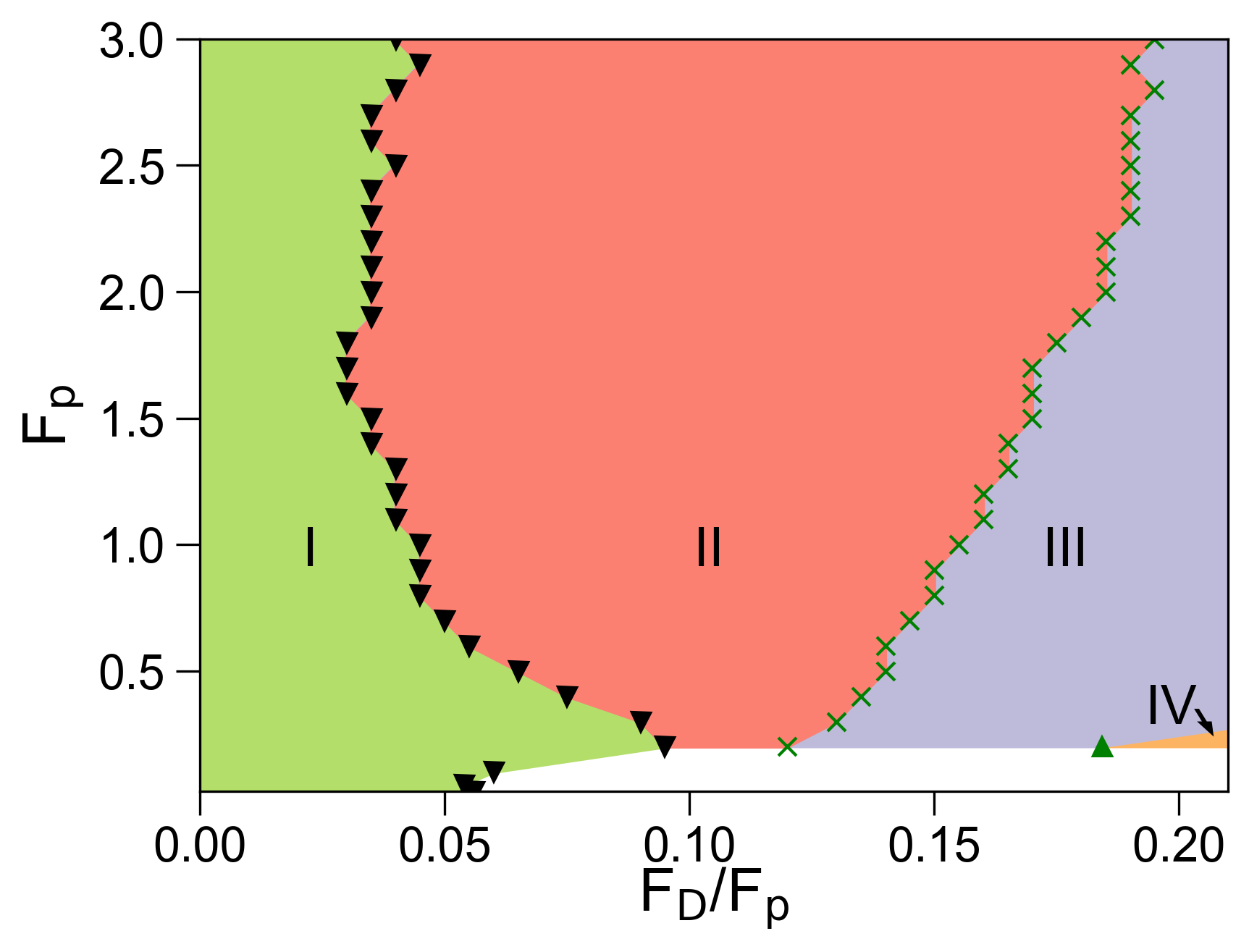}
\caption{Zoomed in version of the phase diagram
  from Fig.~\ref{fig:17} as a function of $F_p$ vs $F_D/F_p$ constructed using
the features identified by PVB PCA.
Green: Pinned (phase I).
Red: Isolated channel flow (phase II).
Purple: Lightly braided channel flow (phase III).
Orange: Heavily braided channel flow (phase IV).
The phase boundaries are obtained from the following measures:
I-II (black down triangles): minimum of $P_1$.
II-III (green crosses): lowest zero crossing of $P_3$.
III-IV (green up triangles): peak of $P_3$.}
\label{fig:PDzoom1}
\end{figure}

\begin{figure}
  \includegraphics[width=\columnwidth]{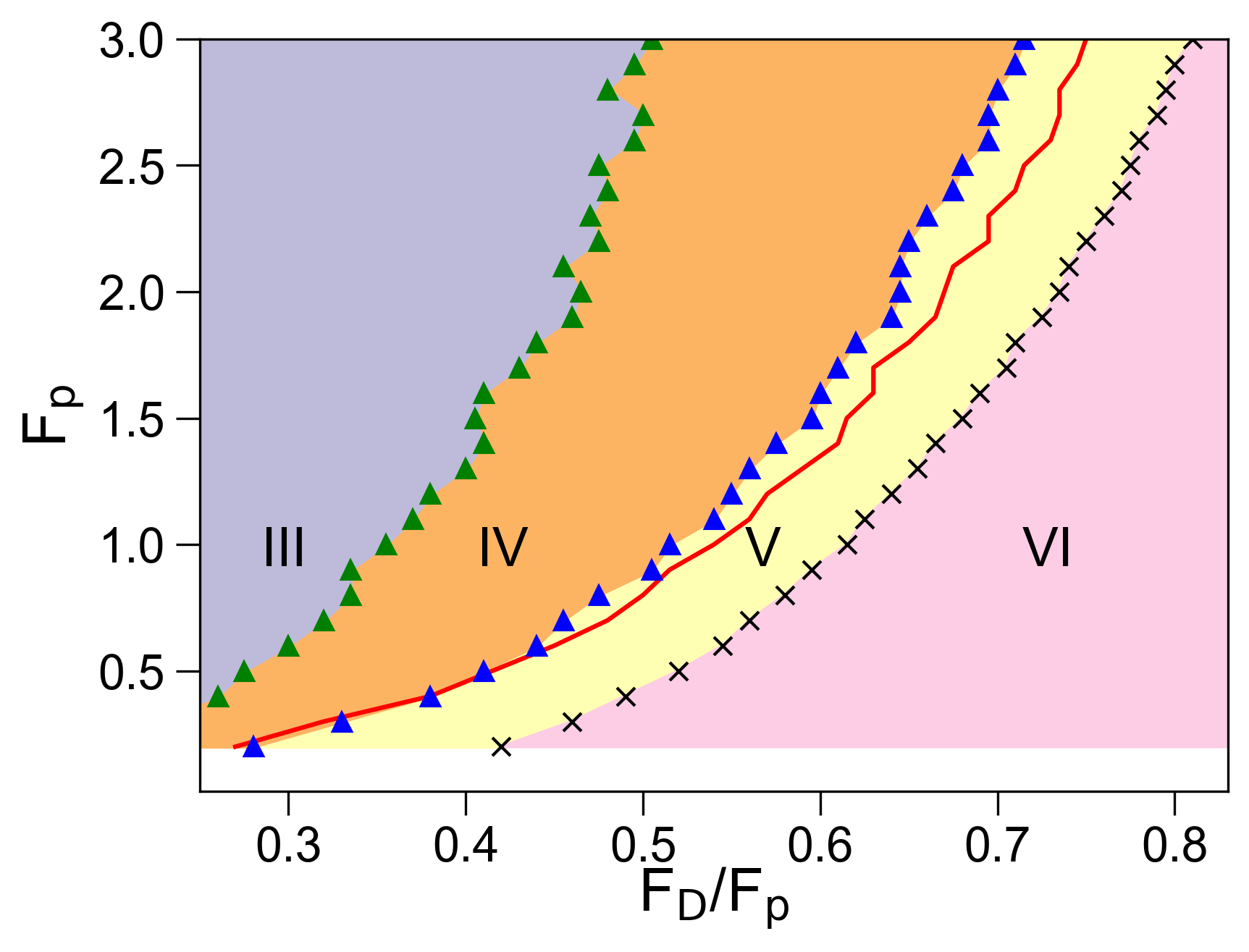}
\caption{
  Zoomed in version of the phase diagram from Fig.~\ref{fig:17}
  as a function of $F_p$ vs $F_D/F_p$ constructed using
the features identified by PVB PCA.
Purple: Lightly braided channel flow (phase III).
Orange: Heavily braided channel flow (phase IV).
Yellow: Inhomogeneous ergodic plastic flow (phase V).
Pink: Emerging one-dimensional flow (phase VI).
The phase boundaries are obtained from the following measures:
III-IV (green up triangles): peak of $P_3$.
IV-V (blue up triangles): peak of $P_2$.
V-VI (black crosses): zero crossing of $P_1$.
The solid red line indicates the peak in $dV/dI$, which does not
directly coincide with the IV-V phase boundary.}
\label{fig:PDzoom2}
\end{figure}

\begin{figure*}
  \includegraphics[width=\textwidth]{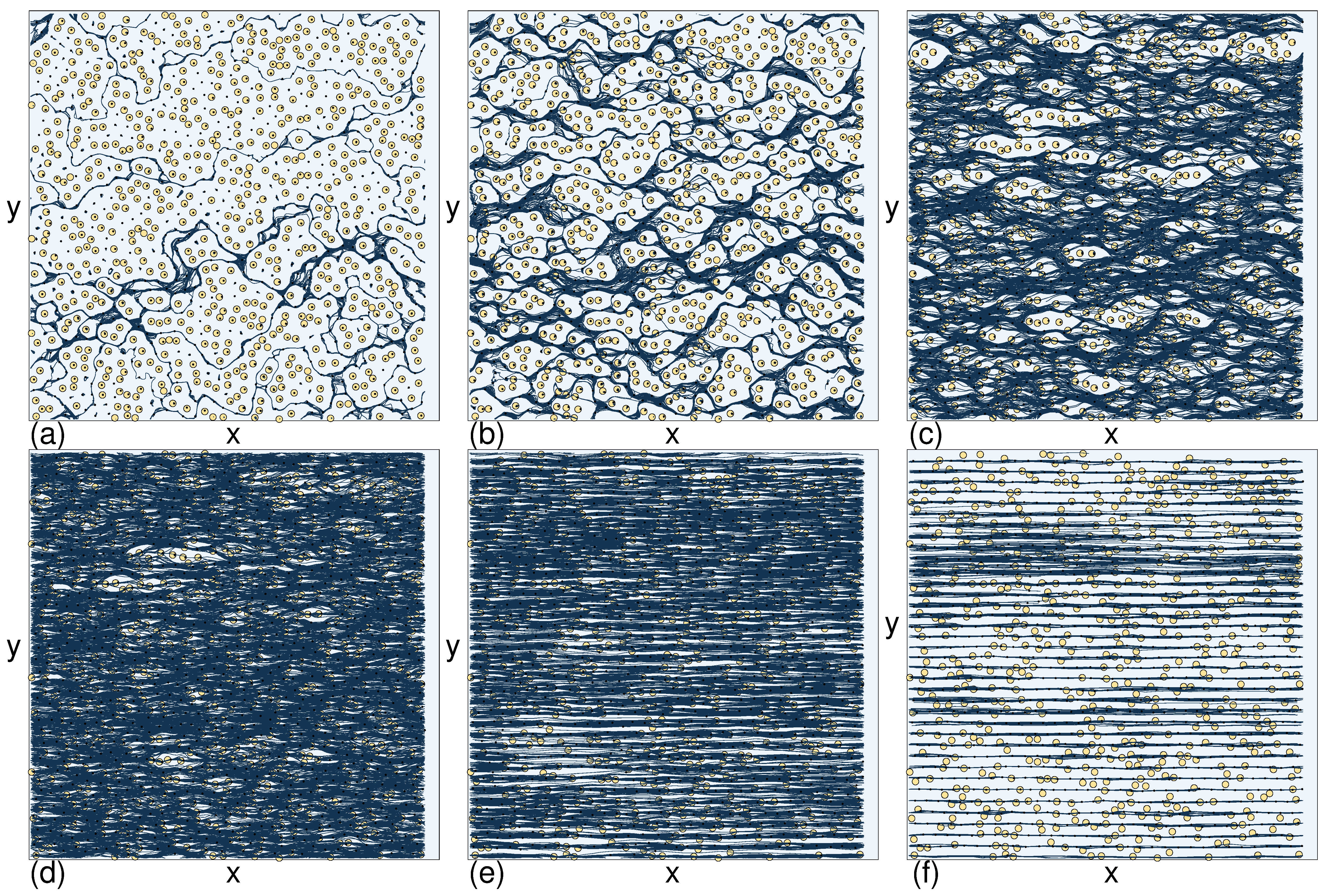}
\caption{Images of the pinning sites (open circles),
vortex positions (dots), and vortex trajectories (lines)
in the different dynamic flow states taken from near the
centers of the phases in the $F_p=1.5$ sample. 
In each panel, the amount of time over which the trails are drawn is
equal to the amount of time that it would take an individual vortex
traveling through a pin-free sample to move a distance of 85$\lambda$
when subjected to the driving force $F_D$.
(a) The isolated channel flow phase II at $F_D/F_p=0.1$.
(b) The lightly braided channel flow phase III at $F_D/F_p=0.285$.
(c) The heavily braided channel flow phase IV at $F_D/F_p=0.5$.
(d) The inhomogeneous ergodic plastic flow phase V at $F_D/F_p=0.63$.
(e) The emerging one-dimensional flow phase VI at $F_D/F_p=1.38$.
(f) The dynamically reordered smectic flow phase VII at $F_D/F_p=2.025$.
}
\label{fig:phasetrail}
\end{figure*}

Based on the different features from
the PVB PCA analysis, we construct the phase diagram illustrated
in Fig.~\ref{fig:17} as a function of $F_p$ versus $F_D/F_p$.
We also present zoomed in versions of the same phase diagram
in Figs.~\ref{fig:PDzoom1} and \ref{fig:PDzoom2}.
At low drives we find a pinned state, phase I.
Phase II is the isolated channel flow state,
illustrated in Fig.~\ref{fig:phasetrail}(a)
at $F_p=1.5$ and $F_D/F_p=0.1$. The number of channels increases
with increasing $F_p$, but in each case, the channels remain narrow and
nonbraiding. Channel width increases somewhat with increasing $F_p$ for
the reasons described above.
The depinning boundary I-II is marked by the location of the minimum in $P_1$.
It is very easily detected and produces a minimum in $P_1$, $P_2$, and
$P_3$, but can also be observed in the velocity-force curve using
the traditionally measured signature of the onset of
a finite velocity.
A more detailed view of the structure of the depinning line
appears in the zoomed in version of the phase diagram
in Fig.~\ref{fig:PDzoom1}.
Phase III is a lightly braided channel flow, shown
in Fig.~\ref{fig:phasetrail}(b) at $F_p=1.5$ and $F_D/F_p=0.285$.
In this phase, the channels develop patches along which the flow can follow
multiple possible pathways.
In between these patches are two-dimensional islands of vortices that remain
permanently pinned.
The II-III boundary is not detected
by any traditional measures of which we are aware.
It is marked by the lowest zero crossing of $P_3$ and is shown in
greater detail in Fig.~\ref{fig:PDzoom1}.
There is a zero crossing of $P_2$ near the center of Phase III
which may be associated with a percolation transition of the trails
transverse to the driving direction.
As shown in Fig.~\ref{fig:phasetrail}(c) for $F_p=1.5$ and $F_D/F_p=0.5$,
phase IV is a heavily braided channel flow.
Some small islands of pinned vortices are still present
but nearly all of these islands exhibit a one-dimensional morphology and have
their long axis aligned with the direction of drive.
The III-IV boundary falls close to the minimum of $p_6$; however, this minimum
is fairly shallow and is not well defined.
We find that a much sharper measure of the boundary can be obtained
from the peak of $P_3$, as shown in Fig.~\ref{fig:17} and with
greater detail in Fig.~\ref{fig:PDzoom2}.
Phase V is inhomogeneous ergodic plastic flow,
where the vortices are able
to access all portions of the sample
as illustrated in Fig.~\ref{fig:phasetrail}(d) at $F_p=1.5$ and
$F_D/F_p=0.63$, but there are significant variations
in vortex velocity as a function of both time and space.
No vortices are permanently pinned; however, some vortices can be
temporarily pinned. 
The IV-V boundary is identified by the peak of $P_2$,
which also coincides with the second zero crossing of $P_3$, and it
has not been observed with any previous measures.
Although this boundary is in
the general vicinity of the peak in $dV/dI$,
indicated by the solid line in Figs.~\ref{fig:17} and \ref{fig:PDzoom2}, 
it has a distinctly different shape, falling slightly
above the peak in $dV/dI$ for low values of $F_p$, and 
well below the $dV/dI$ peak at high values of $F_p$.
In our past work, we have not been able to identify a dynamic phase
transition that is directly associated with the $dV/dI$ peak.
We believe that this is because the $dV/dI$ peak
involves a convolution of the number of moving vortices and their
average velocity.
In Phase VI, shown in Fig.~\ref{fig:phasetrail}(e) at $F_p=1.5$ and
$F_D/F_p=1.38$, one-dimensional channels of flow are beginning to
emerge. Vortices are no longer able to diffuse
freely perpendicular to the driving direction, but are becoming
confined to channels that grow better defined as $F_D$ increases.
The V-VI boundary is determined by the zero crossing of $P_1$,
which coincides with the upper minimum of $P_3$. We note
that $P_1$ provides us with only two signatures: a dip at depinning and
a zero crossing at the V-VI boundary.
During phase VII, the remaining dislocations in the lattice reduce
in density to produce a well-formed dynamically reordered
smectic state. The vortices are now very closely confined to one-dimensional
channels in nearly all of the sample, as illustrated in
Fig.~\ref{fig:phasetrail}(f) at $F_p=1.5$ and $F_D/F_p=2.025$.
The VI-VII boundary is marked by the upper zero crossing of $P_2$.
We note that the third zero crossing of $P_3$ lies a distance of about
$F_D/F_p=0.25$ above the VI-VII boundary but becomes poorly defined
for $F_p<0.5$.

\begin{figure}
  \includegraphics[width=\columnwidth]{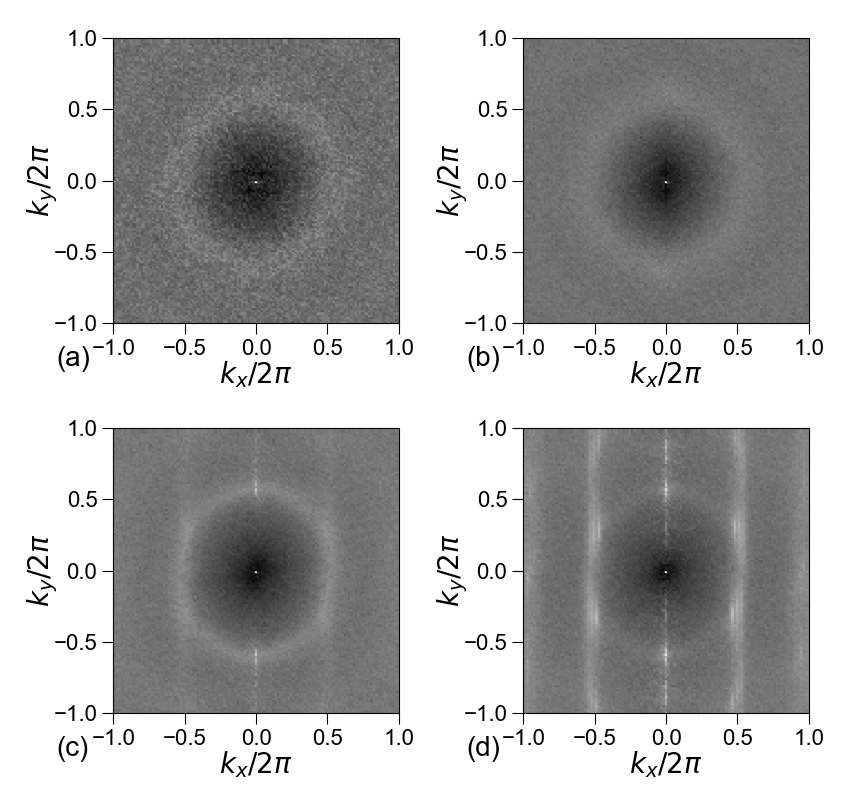}
\caption{Height field plots of the structure factor $S(q)$ as a function of
  $k_x/2\pi$ versus $k_y/2\pi$ for the $F_p=1.5$ sample from
  Fig.~\ref{fig:phasetrail}.
  (a) Liquid structure in the lightly braided channel flow phase II at
  $F_D/F_p=0.1$.
  (b) Liquid structure in the inhomogeneous ergodic plastic flow phase V
  at $F_D/F_p=0.5$.
  (c) Onset of smectic ordering in the emerging one-dimensional flow phase VI
  at $F_D/F_p=1.38$.
  (d) Smectic ordering in the dynamically reordered smectic flow phase VII
  at $F_D/F_p=2.025$.
}
\label{fig:sofq}
\end{figure}

\begin{figure}
  \includegraphics[width=\columnwidth]{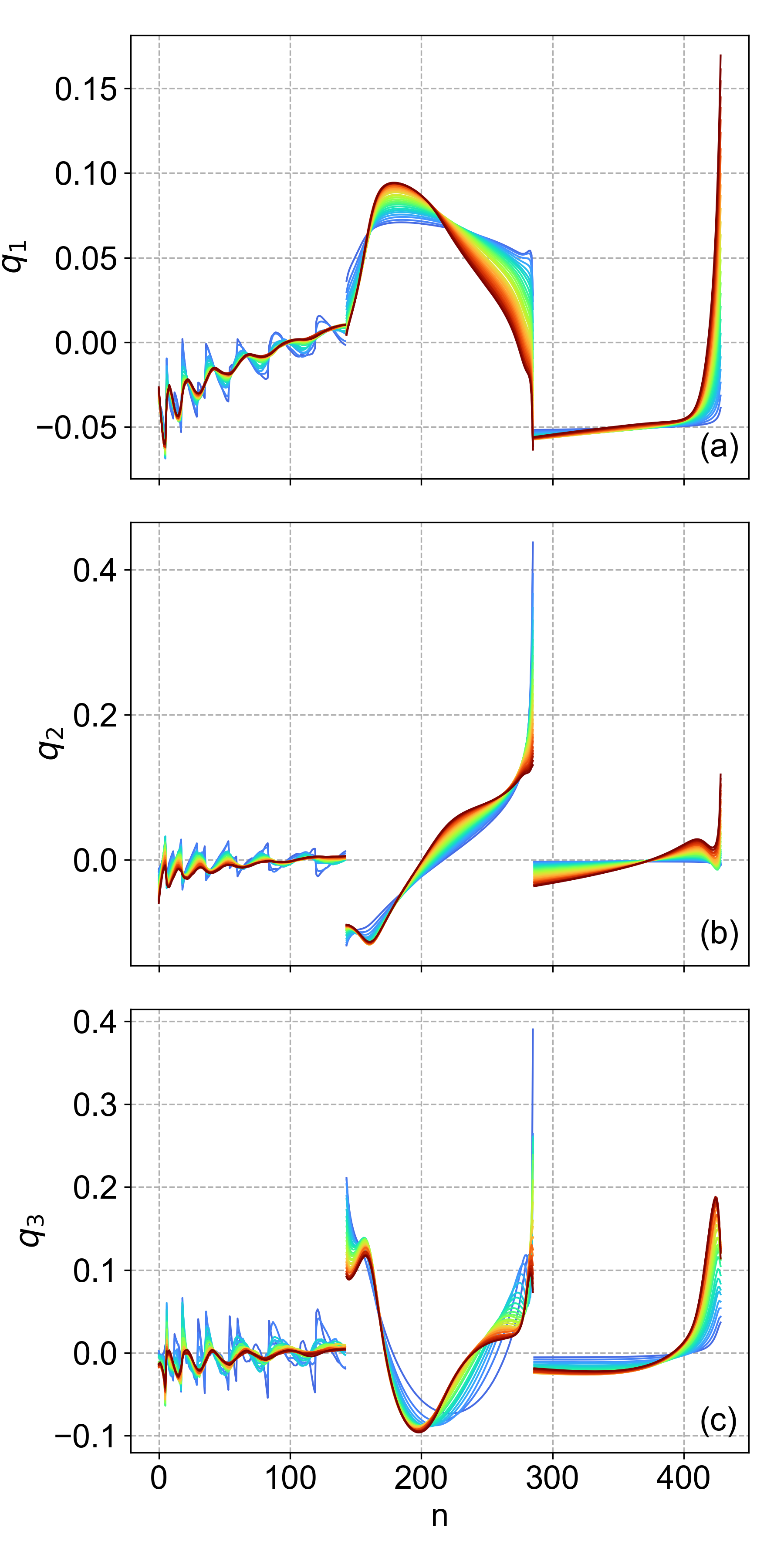}
\caption{The first three rows of the transformation matrix
  matrix ${\bf W}$ vs component number $k$ for the same system from Fig.~\ref{fig:1}
at $F_{p}= 0.2$ (dark blue)
0.3, 0.4, 0.5, 0.6, 0.7, 0.8, 0.9, 1.0, 1.1, 1.2, 1.3, 1.4, 1.5, 1.6,
1.7, 1.8, 1.9, 2.0, 2.1, 2.2, 2.3, 2.4, 2.5, 2.6, 2.7, 2.8, 2.9,
and 3.0 (dark red).
The first 144 entries of $k$
correspond to the average distance $\bar{r}_n$ to the 144 closest
neighbors of the probe particles, sorted in order from smallest to
largest distance.
The second 144 entries of $k$ correspond to the $x$ velocity
difference $\bar{v}_{x,n}$ with those same 144 closest neighbors,
separately sorted in order from smallest to largest velocity difference.
The final 144 entries of $k$ correspond to the $y$ velocity
difference $\bar{v}_{y,n}$ with those same 144 closest neighbors,
again separately sorted in order from smallest to largest velocity difference.
(a) $q_1$, the first row of the transformation matrix, which impacts $P_1$.
(a) $q_2$, the second row of the transformation matrix, which impacts $P_2$.
(a) $q_3$, the third row of the transformation matrix, which impacts $P_3$.
}
\label{fig:q123}
\end{figure}

\section{Discussion}

In our system, it is possible that one or more of the transitions
we observe
are true nonequilibrium dynamic phase transitions
falling in the directed percolation class
\cite{Fisher98,Hinrichsen00,Fily10,Reichhardt17}.
Directed percolation and conserved directed percolation
often describe nonequilibrium phase transitions \cite{Hinrichsen00},
and have been observed for random organization
\cite{Corte08,Hexner15},
plastic flow at yielding transitions \cite{Nagamanasa14}, and
transitions between different turbulent states \cite{Takeuchi07}.
There is also evidence that
the plastic depinning
transition falls into the class of directed percolation
\cite{Reichhardt09,Okuma11,PerezDaroca11,Okuma12,Shaw12,Bermudez20}.
Studies suggest that the transition
from a moving liquid to the moving smectic
state in driven vortex systems is
also in the class of directed percolation
\cite{Reichhardt22a,Maegochi22a},
while the transition to a moving lattice state
is a crossover that is better described by a coarsening mechanism
\cite{Reichhardt22a}.
Our driven vortex system exhibits
an extended region of non-ergodic plastic flow, in which
some vortices remain permanently pinned.
The question is whether the
transition from non-ergodic to ergodic plastic flow is a true transition,
possibly in the directed percolation class, or if it is some other
type of crossover.
Directed percolation transitions are also
known as absorbing phase transitions, and occur between a state
that is strongly fluctuating and a state that is
dynamically frozen \cite{Hinrichsen00}.
In the case of reversible to irreversible
transitions observed in cyclically driven systems \cite{Corte08,Reichhardt23a},
the reversible phase takes on the role of the
dynamically frozen phase, and the irreversible phase
serves as the fluctuating phase.
Once a system becomes trapped in the reversible phase,
it dynamically freezes.
In the depinning system, the plastic flow phase can be
viewed as the fluctuating phase, while
the pinned phase is the absorbing phase.
Similarly, from the fluctuating plastic flow phase,
it is possible to move into the moving smectic state, which serves
as an effectively dynamically frozen phase.

It is interesting to ask whether the structure detected by PVB PCA
produces any signatures in the structure factor $S(q)$. We illustrate
representative structure factor calculations from the $F_p=1.5$
system in Fig.~\ref{fig:sofq} in phases II, V, VI, and VII.
We find nondescript liquid ordering in all of the plastic flow
phases II through V. The structure factor for phases III and IV are not
shown because they have no features to distinguish them from
the phase V state shown in Fig.~\ref{fig:sofq}(b). There are two main
drawbacks of using the structure factor to detect different
dynamic flow states. The first is that, as an inverse space measure,
$S(q)$ smears out local variations in vortex structure in favor of
global changes in vortex density and arrangement. This prevents
$S(q)$ from detecting the presence or absence of isolated patches
of pinned vortices. The second limitation is that the standard $S(q)$
does not incorporate velocity but relies only on the positions of
the vortices. Although it is possible to measure a dynamical
structure factor, this introduces the complication of frequency
dependence and also is still not able to resolve local variations in
velocity such as those found in the braided channel flow states.

Figure~\ref{fig:phasetrail} shows clear images of the different dynamical
flow states, but constructing such a figure without the guidance of the
PVB PCA analysis is prohibitively difficult. Each figure panel requires
high resolution data on the vortex trajectories, and although it is not
impossible to collect data at this resolution throughout the entire
simulation, producing and viewing an image of the trajectories for every
applied current or even for a large number of sampled currents is
beyond the capability of even the most dedicated graduate student.
The system exhibits crossovers rather than sharp transitions between
the plastic flow phases II through V, and the likelihood of capturing
a good representative example of one of these flow states by undirected
sampling is very small. With the phase boundaries from the PVB PCA, we
were able in this study to go directly to individual currents of interest
and run
a high resolution simulation at a single current, starting from an
individual frame of the lower resolution movie used to construct the
PCA feature vectors. We note that it is not possible to
get an accurate picture of the dynamic phases by suddenly applying a nonzero
current to the original zero current starting state of the vortex lattice
due to the long transient times associated with the plastic depinning
transition, which falls in the directed percolation universality class
\cite{Reichhardt09}.

One of the key advantages of the PCA approach is that it is possible to
use the feature vector to combine many different types of variables, such
as particle locations and velocities, and produce individual order
parameters $P_n$ where the peaks, dips, and zero crossings are
correlated with changes in the structure and dynamics of the system.
The disadvantage of PCA compared to traditional measures
such as $p_6$ or spatial correlations
it that the physical meaning of $P_n$ is not immediately obvious; however,
the $P_n$ measurements can provide valuable guides to the locations in
parameter space in which dynamical changes are likely to be occurring,
making it possible to narrow down the window to explore more carefully with
traditional physically motivated measurements.
It is possible to gain some insight into what PCA has detected by
inspecting individual rows of the transformation matrix ${\bf W}$,
as shown for the first three rows $q_1$, $q_2$, and $q_3$ versus
component number $k$ in Fig.~\ref{fig:q123}.
These rows affect the values obtained for $P_1$, $P_2$, and $P_3$,
respectively. Each of the three PCA-determined order parameters
involves a distinct relative weighting of the position information and
relative velocity information. The first principal component has heavy
weighting of the largest $y$ velocity difference and the structure of
the nearest neighbor distances. The second principal component puts
a significant amount of emphasis on the largest and smallest $x$
velocity differences. The third principal component has a more
even weighting of the three elements that have been combined in
constructing each feature vector.

In the case of dynamical reordering or dynamical freezing of a moving
lattice, counting the fraction of defects or measuring the structure factor
can provide good order parameters; however, crossovers or transitions
among different disordered states are much more difficult to
distinguish, and here PCA can provide well defined information about where
to look for a disorder-disorder transition.
The same PVB PCA approach could be applied to other types of
depinning transitions such as interface depinning or the plastic
motion that occurs during the yielding process.
Other systems could also be considered, such as the melting and freezing
of a non-driven state in the presence of quenched disorder, where the
system can transition from a thermal fluid to a pinned glass. PVB PCA
in which
the absolute velocity rather than velocity components are considered
could be a good method for detecting signatures of multiple melting
regimes.
There has also been work on vortices driven over quenched disorder under
conditions where thermal effects are important, and in this case,
the drive at which dynamical reordering occurs diverges at the melting
temperature of the quenched disorder-free system
\cite{Bhattacharya93,Koshelev94}.
A future direction could be to apply PVB PCA to these systems to determine
whether PCA can distinguish thermal fluids from the 
$T = 0.0$ moving fluid we consider in the present work.
Other systems to consider are those with avalanche-type dynamics and
quenched disorder, to see whether PVB PCA can distinguish between
plastic and elastic avalanche events. PCA could also be applied to
depinning and sliding transitions in systems with 
periodic pinning arrays,
where there can be different types of kink or antikink flows
as well as disordered to laminar flow transitions
\cite{Gutierrez09,Bohlein12,Reichhardt17},
and permit a comparison between the results in the periodic pinning system and
the random pinning results we consider here.

We performed our study for superconducting
vortex pinning in a 2D system,
but there are also a variety of systems
in which 3D vortex depinning and driven plastic flow phases
can occur, such as samples in which the vortices can be
represented as
elastic lines that depin \cite{Reichhardt17}, or
systems where the vortex lines can break apart and undergo
3D to 2D transitions associated with different types of
plastic flow \cite{Kolton00,Olson01a}.
The PVB PCA approach could be generalized and applied to such
higher dimensional systems.

\section{Summary}

We have examined the depinning and nonequilibrium
flow phases for driven vortices in two-dimensional systems
with random quenched disorder, where we correlate  measures
such as the density of topological defects and
the velocity-force curves with order parameters extracted from
a principal component analysis (PCA).
We develop particle-velocity based (PVB) feature vectors for the
PCA approach that employ
the vortex positions and information about the vortex velocity
components both parallel and perpendicular to the applied drive.
For strong disorder or high pinning force,
the system depins plastically, and there is a dynamic reordering transition
at high drives into a moving smectic.
The fraction of six-fold coordinated vortices passes through a minimum
in the plastic flow regime before saturating at a higher value for
large drives,
and the $dV/dF_{D}$ curves have an initial slow linear increase followed by
a peak and an asymptotic approach to $dV/dF_D=1.0$ at high drives.
We show that the order parameters
$P_1$, $P_{2}$, and $P_{3}$ derived from PVB PCA have local maxima, minima,
and zero crossings that are
correlated with signatures in the transport and structural measurements,
enabling PCA to identify
the depinning transition point
and the dynamical ordering transition.
The PCA measures also provide evidence that when the pinning is
sufficiently strong, there are
multiple distinct plastic flow states. In non-ergodic plastic flow, the motion
occurs in well-defined static channels that visit only a portion of
the total sample, while the remainder of the vortices in the
sample remain permanently pinned.
In the ergodic plastic flow state,
vortices are at most temporarily pinned, 
and flow can occur everywhere in the sample.
The transition from non-ergodic to ergodic plastic flow
detected by PCA occurs 
near but not at the peak in $dV/dF_{D}$.
The PCA measures also provide evidence 
of a crossover within the non-ergodic
plastic flow regime from a weakly interacting
channel state to a
strongly intersecting channel regime.
In the ergodic plastic flow regime, PCA indicates that
there is a crossover from a regime where some vortices are temporarily
pinned to a disordered flowing state in which the vortex velocity
distribution function is still bimodal but none of the vortices are
pinned even temporarily.
Performing the PVB PCA analysis on both the elastic and plastic depinning
regimes, we find very different signatures in the order parameters.
For elastic depinning, there is only a feature in $P_n$ at
the depinning transition.
In contrast, in the plastic depinning systems
there are multiple maxima, minima, and zero crossings in $P_n$,
indicating the presence of multiple distinct flow regimes.
Based on the signatures extracted from
PVB PCA, we propose a new dynamical phase diagram
for driven vortices
containing both elastic and plastic depinning regimes,
multiple non-ergodic plastic flow states including isolated channel
flow, lightly braided channel flow, and heavily braided channel flow,
an ergodic plastic flow phase, an emerging smectic phase, and a fully
developed moving smectic state.
The boundaries between the lightly and heavily braided channel flow
states and between the heavily braided and ergodic plastic flow phases
have not been identified previously with traditional measures to our
knowledge.
Our results suggest that PVB PCA can serve as a useful analysis
tool for various other systems that exhibit nonequilibrium phase
transitions.

\acknowledgments
We gratefully acknowledge the support of the U.S. Department of
Energy through the LANL/LDRD program for this work.
  This work was supported by the US Department of Energy through
  the Los Alamos National Laboratory.  Los Alamos National Laboratory is
  operated by Triad National Security, LLC, for the National Nuclear Security
  Administration of the U. S. Department of Energy (Contract No. 892333218NCA000001).
  
\bibliography{mybib}

\end{document}